%% file: submission.tex
\documentclass[10pt,twocolumn,letterpaper]{article}

\usepackage{iccv}
\usepackage{times}
\usepackage{epsfig}
\usepackage{graphicx}
\usepackage{amsmath}
\usepackage{amssymb}

\usepackage{enumitem}
\usepackage{multirow}
\usepackage{booktabs}
\usepackage{bbm}
\usepackage{arydshln}
\usepackage{xspace}
\usepackage{cite}
\usepackage{adjustbox}
\usepackage[pagebackref=true,breaklinks=true,letterpaper=true,colorlinks,bookmarks=false]{hyperref}

\renewcommand{\etal}{\textit{et al}.\xspace}
\renewcommand{\ie}{\textit{i}.\textit{e}.\xspace}
\renewcommand{\eg}{\textit{e}.\textit{g}.\xspace}

\newcommand{\x}{\mathbf{x}}
\newcommand{\y}{\mathbf{y}}
\newcommand{\z}{\mathbf{z}}
\newcommand{\m}{\mathbf{m}}
\newcommand{\w}{\mathbf{w}}

\DeclareMathOperator*{\argmin}{arg\,min}
\iccvfinalcopy % *** Uncomment this line for the final submission

\def\iccvPaperID{5705} % *** Enter the ICCV Paper ID here

\ificcvfinal\fi

\begin{document}

%%%%%%%%% TITLE
\title{Variable-Rate Deep Image Compression through \\ Spatially-Adaptive Feature Transform}

\author{Myungseo Song \hspace{1.0cm} Jinyoung Choi \hspace{1.0cm} Bohyung Han\\
ECE \& ASRI, Seoul National University, Korea\\
{\tt\small \{micmic123, jin0.choi, bhhan\}@snu.ac.kr}
}
\maketitle
% Remove page # from the first page of camera-ready.
\ificcvfinal\thispagestyle{empty}\fi

%%%%%%%%% ABSTRACT
%-------------------------------------------------------------------
%		Abstract
%-------------------------------------------------------------------
\input{./sections/Abstract}

%%%%%%%%% BODY TEXT
%-------------------------------------------------------------------
%		Introduction
%-------------------------------------------------------------------
\input{./sections/Introduction}

%-------------------------------------------------------------------
%		Related work
%-------------------------------------------------------------------
\input{./sections/Related_work}

%-------------------------------------------------------------------
%		Methodology
%-------------------------------------------------------------------
\input{./sections/Method}

%-------------------------------------------------------------------
%		Experiments
%-------------------------------------------------------------------
\input{./sections/Experiments}

%-------------------------------------------------------------------
%		Conclusion
%-------------------------------------------------------------------
\input{./sections/Conclusion}

{\small
\bibliographystyle{ieee_fullname}
\bibliography{egbib}
}

% -------------------------------------------------------------------
% 		Appendix
% -------------------------------------------------------------------
\clearpage
\input{./sections/Appendix.tex}

\end{document}

%% file: sections/Abstract.tex
% !TEX root = ./../submission.tex

\begin{abstract}
We propose a versatile deep image compression network based on Spatial Feature Transform (SFT)~\cite{wang2018recovering}, which takes a source image and a corresponding quality map as inputs and produce a compressed image with variable rates.
Our model covers a wide range of compression rates using a single model, which is controlled by arbitrary pixel-wise quality maps.
In addition, the proposed framework allows us to perform task-aware image compressions for various tasks, \eg, classification, by efficiently estimating optimized quality maps specific to target tasks for our encoding network.
This is even possible with a pretrained network without learning separate models for individual tasks.
Our algorithm achieves outstanding rate-distortion trade-off compared to the approaches based on multiple models that are optimized separately for several different target rates. 
At the same level of compression, the proposed approach successfully improves performance on image classification and text region quality preservation via task-aware quality map estimation without additional model training.
The code is available at the project website\footnote{\url{https://github.com/micmic123/QmapCompression}}.
\end{abstract}

%% file: sections/Introduction.tex
% !TEX root = ./../submission.tex

\section{Introduction}
\label{sec:introduction}
%-------------------------------------------------------------------------

\begin{figure}[t]
  \centering
  \begin{tabular}{c}
    \includegraphics[width=0.45\textwidth]{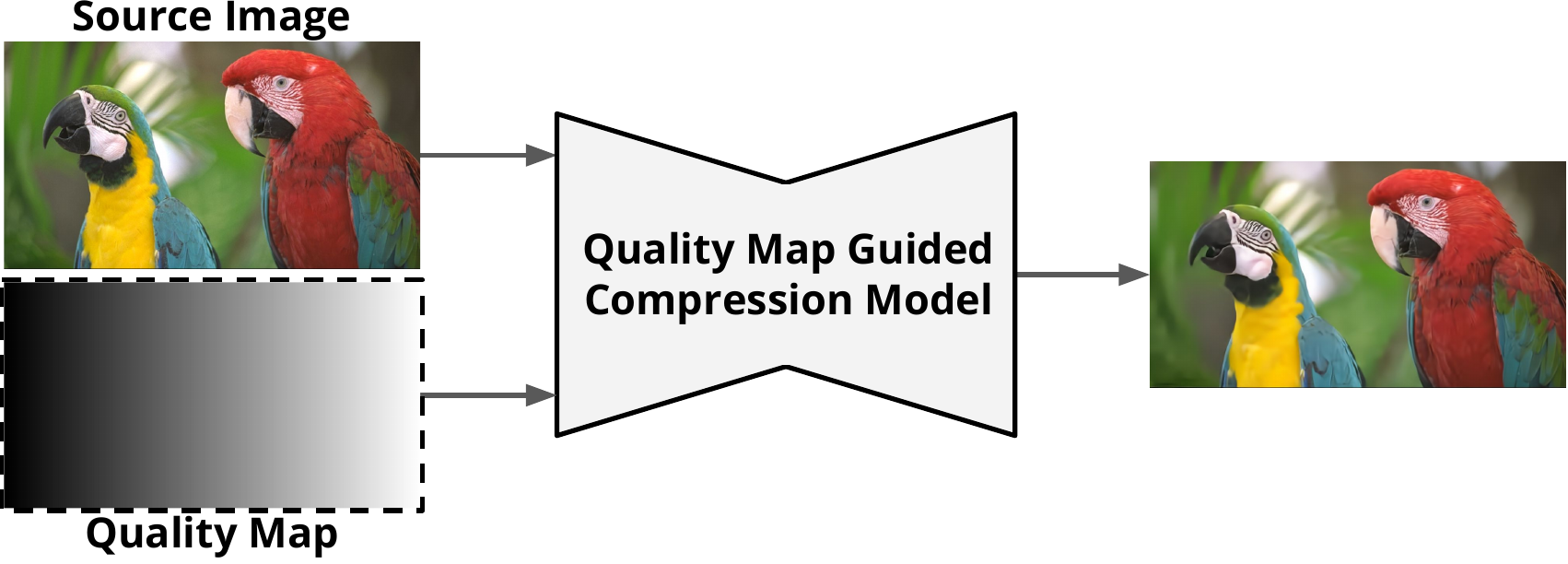} \\
    (a) \\
    \includegraphics[width=0.45\textwidth]{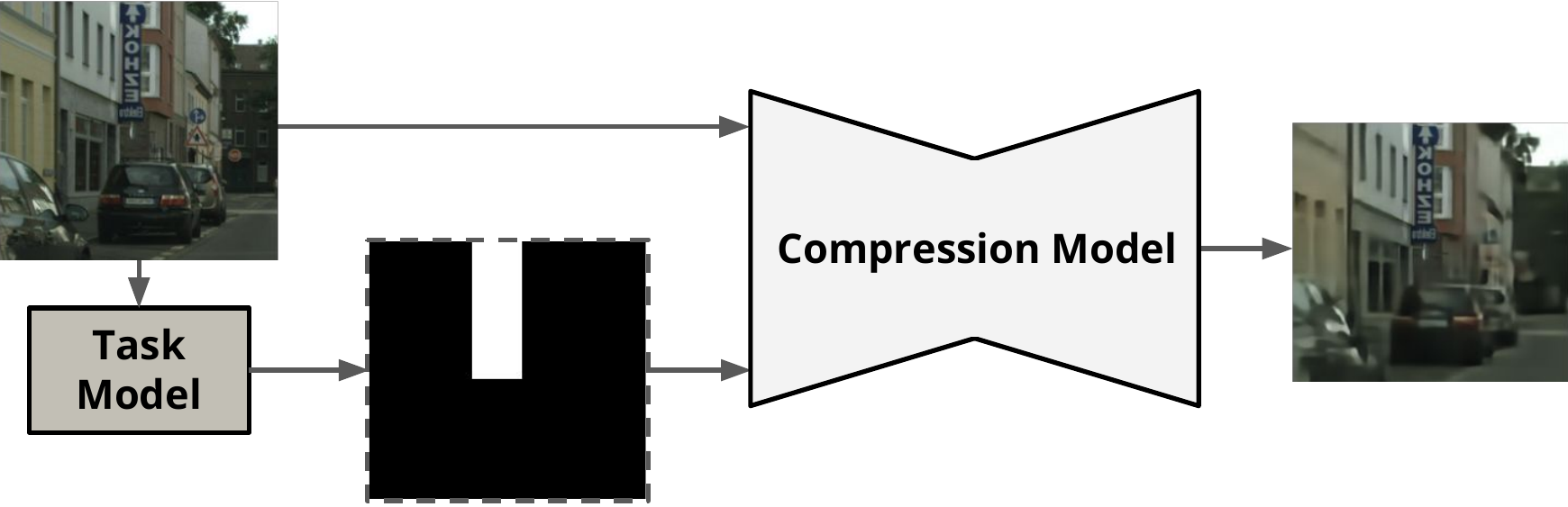} \\
    (b) \\
    \includegraphics[width=0.45\textwidth]{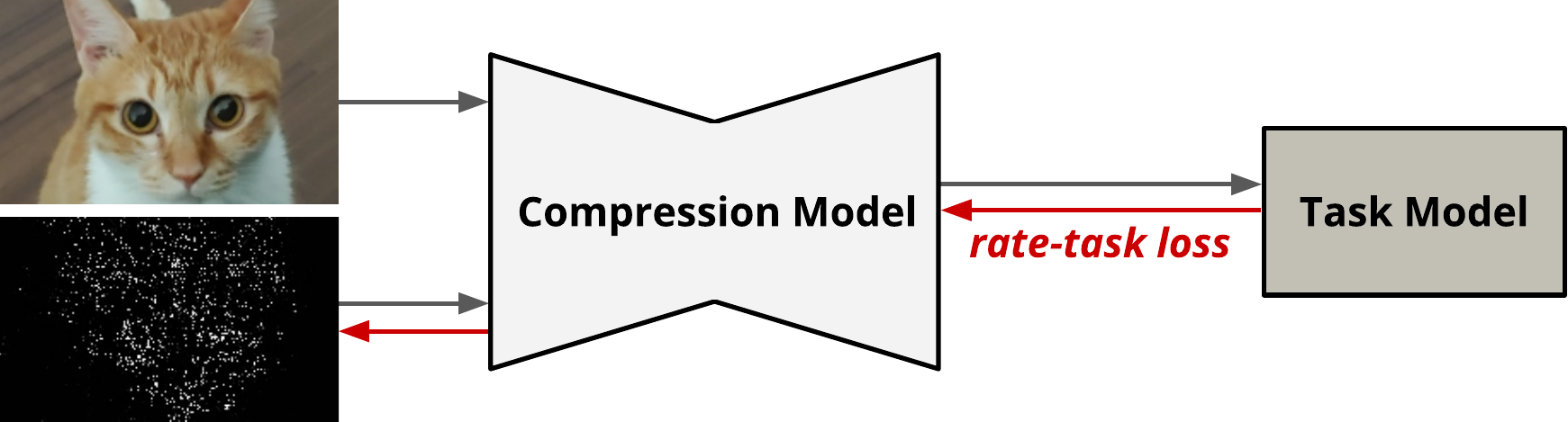} \\
    (c) \\
  \end{tabular}
  \caption{The proposed framework. 
          (a) Our compression model takes a quality map as an input along with an image. 
          The compression for the image is conditioned on the quality map, which indicates pixel-wise importance.
          (b) Task-aware compression at test time via non-uniform quality map.
          One can  create a quality map manually or use an output of a pretrained task model (\eg, object detection results).
          (c) A task-aware quality map from pretrained models can be estimated at encoding phase by minimizing rate-task loss for the quality map without fine-tuning the models.
          }
          \vspace{-0.3cm}
  \label{fig:framework}
\end{figure}

\begin{figure*}[t]
  \centering
  \includegraphics[width=0.96\textwidth]{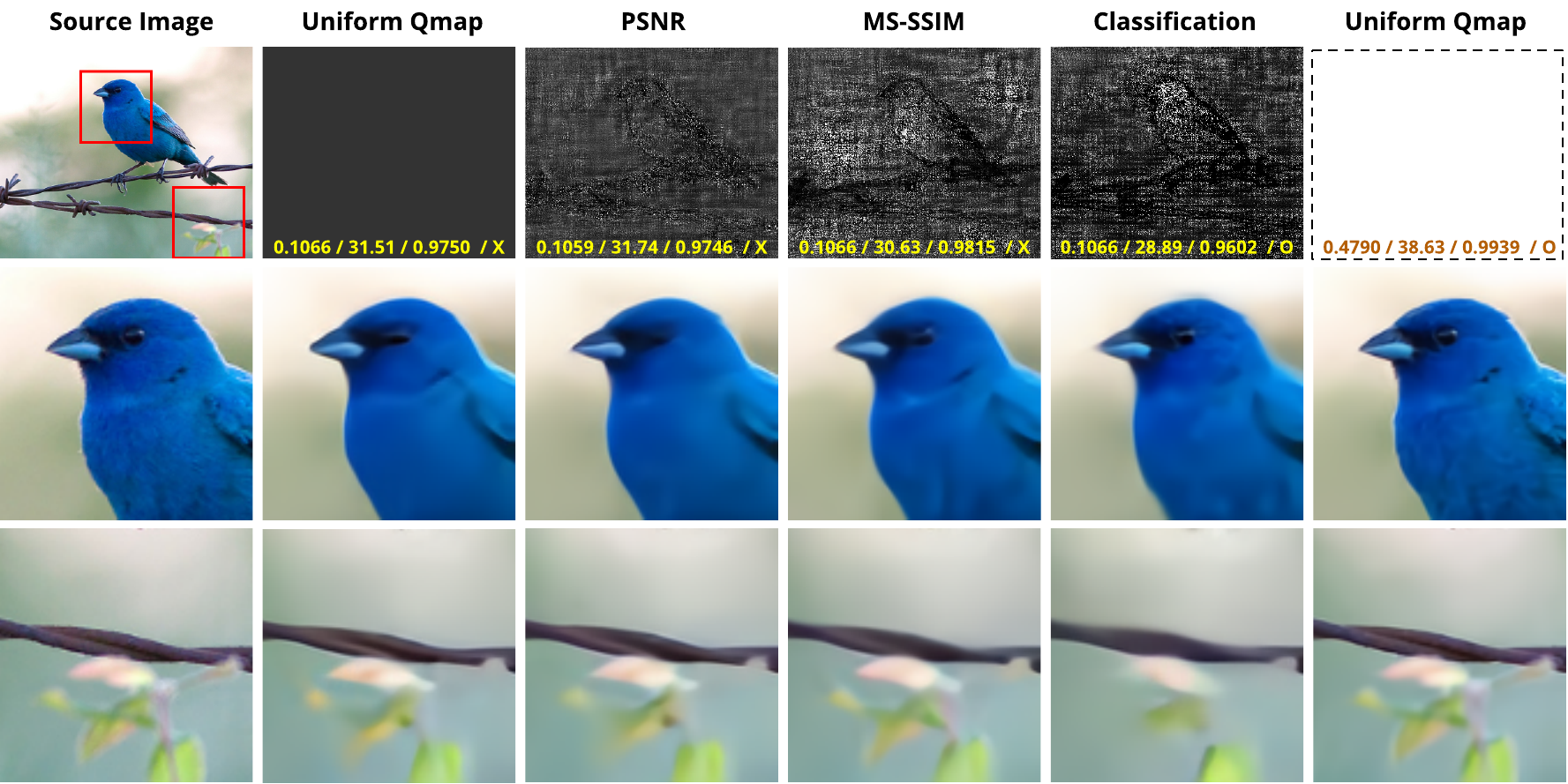}
    \vspace{0.2cm}
  \caption{Compression results with various quality maps using our model. 
  All images are obtained by the same model. 
  Except for uniform quality maps (2\textsuperscript{nd} and 6\textsuperscript{th} column), the quality maps are inferred by optimizing the rate-task loss. 
  The numbers in each quality map denote bits per pixel (bpp)/PSNR (dB)/MS-SSIM/classification result of the corresponding reconstructed image.
  Results by two different uniform maps show that our model adapts well at a wide range of bitrate.
  The rest of the results show that our model can be adjusted to desired task without additional training of the model. 
  For example, one with classification-aware quality map (5\textsuperscript{th} column) only succeeds in classification among all reconstructed images with similar bpp, while having the worst PSNR and MS-SSIM.
  Qualitatively, it maintains the quality of discriminative regions for bird recognition, \eg, an eye, at the expense of the quality in other regions.
  }
  \label{fig:test_image}
\end{figure*}

Image compression has been studied actively for decades and has recently become more critical with exploding use of mobile devices for capturing and sharing images.
Lossy image compression is a particularly useful technique in this trend since it reduces required space and transmission cost significantly at the expense of quality of reconstructed images.
The techniques based on deep learning~\cite{balle2017end, mentzer2018conditional, balle2018variational, minnen2018joint, lee2018context, li2018learning, li2020learning, minnen2020channel, zhong2020channel, mentzer2020high, yang2020variable, choi2019variable, johnston2018improved, toderici2017full} have started to outperform traditional codecs including JPEG~\cite{pennebaker1992jpeg}, JPEG2000~\cite{skodras2001jpeg} and BPG~\cite{bellard2014bpg}.
Many learning-based approaches adopt autoencoder networks for coding nonlinear transformation and optimizing rate-distortion trade-off~\cite{davisson1972rate}.
They have arisen as new candidates for next generation image compression standards due to their high performance and applicability compared to the traditional hand-engineered codecs.

The rate-distortion optimization in learned lossy image compression methods is realized by minimizing a combined loss function based on compression ratio and distortion between an original and a corresponding output image.
For the objective function, most of existing algorithms rely on a uniform compression ratio over image space.
However, all pixels in an image are not equally important and a spatially-adaptive image compression by identifying regions of interest (ROIs) would be desirable for better performance.
On the other hand, existing models are typically optimized for a single target compression rate and their extensions to multiple rates are not straightforward. 

This paper introduces a variable-rate image compression network based on an importance map with spatially adaptive continuous values.
Specifically, we optimize the objective function pertaining to rate-distortion trade-off, where image distortion is constrained by a 2D real-valued quality map that defines pixel-wise weight for computing mean squared errors (MSEs). 
The proposed approach employs spatially-adaptive affine transform modules, which perform pixel-wise feature transformations and result in compressed images guided by a quality map.
Note that our model enables us to compress images with arbitrary compression rates and obtain compressed images with spatially-varying quality given by quality maps.
In addition to the flexibility of image compression, we also propose a technique to automatically generate task-aware quality maps by backpropagation without retraining the models and construct compressed images optimized for target tasks.

Compared to the existing adaptive image compression techniques based on variable rates and ROIs~\cite{yang2020variable, choi2019variable, cui2020g, johnston2018improved, toderici2017full, agustsson2019generative,cai2019end,akutsu2019end,lohdefink2020focussing}, 
the proposed framework is much more flexible and generalizable.
The variable-rate models~\cite{yang2020variable, choi2019variable, cui2020g} have shown comparable performance with single-rate counterparts.
However, Yang~\etal~\cite{yang2020variable} handle only a few discrete levels of compression using an autoencoder.
Choi~\etal~\cite{choi2019variable} and Cui~\etal~\cite{cui2020g} have proposed continuously variable-rate models, but improper selection of quantization bin size leads to a degradation of rate-distortion performance~\cite{choi2019variable}.
More importantly, all of these methods do not consider explicit spatial importance for image compression.
{A variable-rate approach based on a recurrent neural network (RNN)~\cite{johnston2018improved} evaluates the distortion of individual patches in a source image to compute weighted distortion.
However, it makes the quality of each patch roughly uniform, and realizes spatial adaptiveness by introducing a post-processing for dynamic bit allocation.}
Minnen~\etal~\cite{minnen2017spatially} allows coarse (per-patch) quality variation, but suffers from slow coding speed and low quality compared to recent methods.
On the other hand, there exist ROI-based compression methods that reflect the given pixel-wise importance~\cite{agustsson2019generative,cai2019end,akutsu2019end,lohdefink2020focussing}, but they are limited to taking a binary mask and compressing images with predefined discrete levels, too. 

Figure~\ref{fig:framework} illustrates the proposed framework for training and inference. 
The main contributions of our approach are summarized as follows:

\begin{itemize}[label=$\bullet$]
	\item We propose a variable-rate image compression algorithm conditioned on a real-valued quality map, which guides an efficient bit allocations across pixels within an input image.
	\item We design an effective network architecture based on spatially-adaptive feature transform, which takes advantage of spatial information as a prior, for our conditioned image compression.
	      Our model even excels the compression performance of fixed-rate models in a practical range of bitrates.
	\item We introduce a method to estimate task-specific quality map for image compression at test time. The inferred quality maps are efficient to acquire and effective to achieve good performance in the target task.
\end{itemize}

The rest of the paper is organized as follows. 
Section~\ref{sec:related_work} reviews related literatures on deep image compression and Section~\ref{sec:method} presents the proposed method in detail.
We show experimental results and analysis in Section~\ref{sec:experiments}.

%% file: sections/Related_work.tex
% !TEX root = ../submission.tex

\section{Related work}
\label{sec:related_work}

This section first discusses basic image compression techniques based on deep neural networks, and presents two kinds of adaptive models including variable-rate compression and ROI-guided compression.
Then, we describe techniques about spatially-adaptive affine transform, which is closely related to the main component in our algorithm.

%-------------------------------------------------------------------------
\subsection{Deep image compression}
Deep image compression models learn to minimize distortion between a pair of a source image and a reconstructed image while maximizing the likelihood of the quantized latent representation for low entropy coding cost (bitrate).
The trade-off between the rate and the distortion is controlled by a Lagrange multiplier $\lambda$~\cite{davisson1972rate}, but most of existing works~\cite{balle2017end, ayzik2020deep, agustsson2017soft, rippel2017real, mentzer2018conditional, balle2018variational, minnen2018joint, lee2018context, li2018learning, li2020learning, minnen2020channel, zhong2020channel, tschannen2018deep, mentzer2020high} are limited to learning for a fixed value of $\lambda$ and obtaining only a single point in a rate-distortion curve for each trained model.
Various methods have been proposed to improve the rate-distortion trade-off.
For example, \cite{balle2017end, rippel2017real, minnen2017spatially,mentzer2018conditional, lee2018context, minnen2020channel} incorporate entropy prediction of learned representations based on context during training, and \cite{mentzer2018conditional, li2018learning, li2020learning} employ importance maps internally for dynamic bit allocation of latent representations.
Some approaches introduce additional models for hyper-prior, which provide side information for a conditional entropy model~\cite{balle2018variational, minnen2018joint,lee2018context}.

\subsection{Variable-rate compression}
There exist a handful of image compression approaches to support variable-rates using a single model~\cite{yang2020variable, choi2019variable, cui2020g, johnston2018improved, toderici2017full}.
Early methods~\cite{johnston2018improved, toderici2017full} employ RNNs, where the number of iterations are used to control target rates.
However, their processing time for encoding and decoding increases as image quality gets higher, which makes the methods impractical.
More recent works~\cite{yang2020variable, choi2019variable, cui2020g} adopt multiple Lagrangian multiplier values $\lambda$ to define the loss function and allow trained models to handle multiple rate-distortion trade-offs using feature transformations depending on $\lambda$'s.
Specifically, \cite{yang2020variable} presents an image compression network based on an autoencoder that supports multiple discrete levels of compression rate.
Choi~\etal~\cite{choi2019variable} select quantization bin sizes in multiple discrete levels to approximate real-valued rates.
However, it is not straightforward to determine the discrete levels and the bin sizes, which makes the quality of output images suboptimal.
On the other hand, Cui \etal~\cite{cui2020g} modulates compression rates continuously via interpolation of the learned parameters in a pretrained discrete variable rate model.
All these aforementioned approaches commonly adopt channel-wise affine transform conditioned on Lagrange multipliers and show comparable accuracy with single-rate models trained independently.

\subsection{ROI-based compression}
ROI-based models for image compression take a binary mask as an additional input to maintain higher reconstruction quality in the ROI while ignoring or discounting other areas~\cite{agustsson2019generative,cai2019end,akutsu2019end,lohdefink2020focussing}.
Agustsson~\etal~\cite{agustsson2019generative} synthesizes unimportant regions using a semantic label map using a generative adversarial network (GAN) to achieve extremely low bitrates. 
Another GAN-based method~\cite{lohdefink2020focussing} minimizes the MSE of important regions directly while reducing the distortion of remaining parts indirectly using a feature matching loss for the entire image.
Cai~\etal~\cite{cai2019end} present a similar approach based on MS-SSIM as a distortion metric, where they train the model to predict ROI in a supervised manner.
Akutsu~\etal~\cite{akutsu2019end} employ weighted MS-SSIM conditioned on ROI masks with specific target distortion values.
It is worth noting that, to our knowledge, all existing ROI-based models are limited to using binary masks. 

\subsection{Spatially-adaptive affine transform}
The adaptive feature transformation applies a transformation with affine parameters dynamically generated from an external information.
After the adaptive instance normalization~\cite{huang2017arbitrary} have been proposed for style transfer,
\cite{wang2018recovering, park2019semantic, kim2020transfer} extend this idea to spatially-varying affine transform (denormalization) with element-wise distinction for super-resolution, semantic image synthesis, and denoising.
Unlike \cite{park2019semantic, kim2020transfer}, spatial feature transform (SFT)~\cite{wang2018recovering} modulates intermediate feature maps of a network without normalization.
All these works rely on the external inputs when generating the spatially-varying parameters.
On the other hand, our approach employs the original input of the network in addition to the external information for SFT.

%% file: sections/Method.tex
% !TEX root = ./../submission.tex

%-------------------------------------------------------------------------
\section{Image Compression Guided by Quality Map}
\label{sec:method}
This section presents our main idea of variable-rate image compression via spatially-adaptive feature transform guided by a pixel-wise quality map.

\begin{figure}[t]
  \centering
  \includegraphics[width=0.95\linewidth]{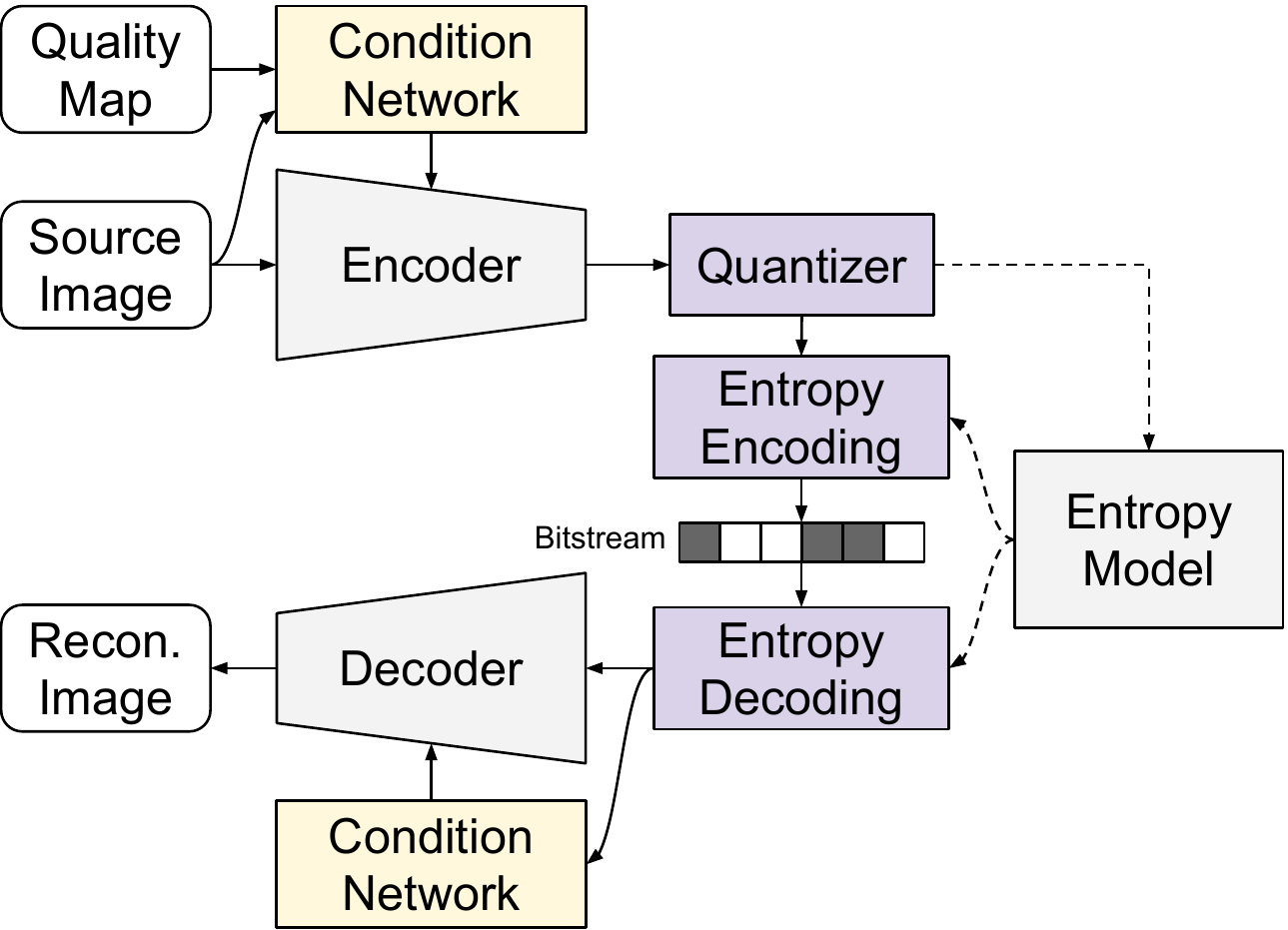} \\
  \vspace{0.2cm}
  \caption{High-level architecture of our model. To perform effective compression conditioned on the quality map, we introduce the condition networks to the compression model.
          }
  \label{fig:architecture_abstraction}
\end{figure}

\subsection{Overview}
The proposed algorithm is a generalized and integrated version of ROI-based and variable-rate image compression methods.
For a source image $\x=[x_i]_{i=1:N}$, our model takes a quality map $\m=[m_i]_{i=1:N}$ $(m_i \in [0, 1])$ as its side information to reflect spatial importance of $\x$.
The quality map $\m$ defines pixel-wise quality levels after compression.
As in the standard image compression approaches, our model consists of three primary components, an encoder, a quantizer, and a decoder.
The encoder transforms $\x$ conditioned on $\m$ to a latent representation $\y$, which is then quantized to $\hat{\y}$ by the quantizer.
Since rounding operation for quantized values is non-differentiable, it is relaxed to a differentiable alternative---additive uniform noise~\cite{balle2017end}---during training.
After the quantization, an entropy coding, \eg, arithmetic coding~\cite{rissanen1981universal}, is performed on $\hat{\y}$ to save it as a lossless bitstream.
For reconstruction, the decoder generates a reconstructed image ${\x}'$ from $\hat{\y}$.

\begin{figure}[t]
  \centering
  \includegraphics[width=1\linewidth]{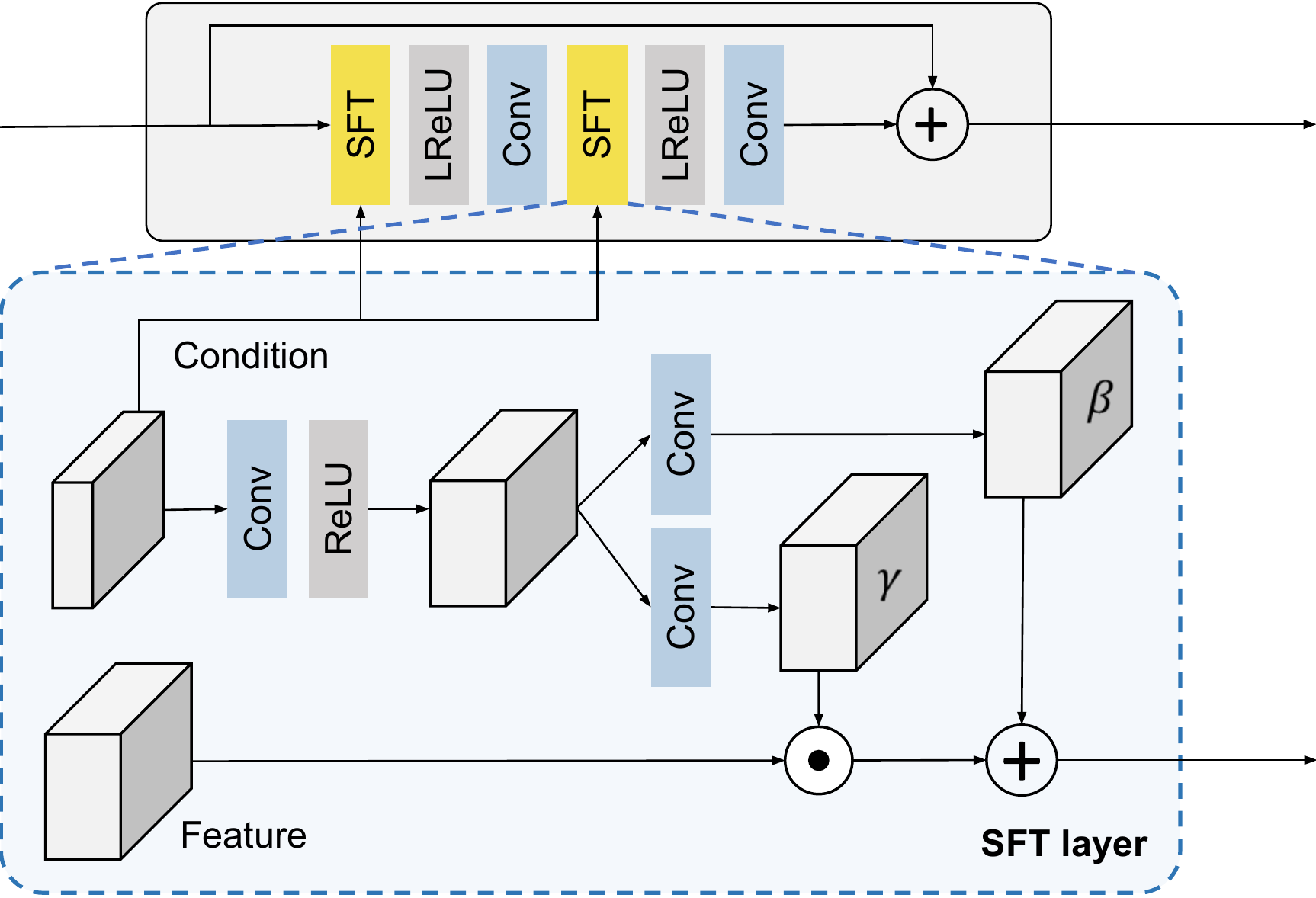} \\
  \vspace{0.2cm}
  \caption{Illustration of Spatial Feature Transform (SFT) layer~\cite{wang2018recovering} and residual block we implement, which are related to \cite{park2019semantic}.
           The SFT layer takes an intermediate feature of previous layer and a prior condition to produce element-wise scaling and shifting parameters $\gamma$ and $\beta$.
           As a result, the element-wise affine transformation is performed to the feature.
          }
  \label{fig:sft}
\end{figure}

Figure~\ref{fig:architecture_abstraction} depicts the high-level concept of the proposed model. 
The unique components in our approach are the condition networks, which are designed to utilize quality maps effectively. 
We will make detailed discussion about the architecture of our model in Section~\ref{sec:net_architecture}.

\subsection{Rate-distortion loss}
The goal of lossy image compression is to minimize the length of the bitstream and the distortion between $\x$ and ${\x}'$ concurrently.
This objective raises an optimization problem of minimizing $R+\lambda D$, where a Lagrange multiplier $\lambda$ of a fixed value determines the trade-off between the rate $R$ and the distortion $D$.
On the other hand, our model achieves the variable-rate compression by minimizing $R+\Lambda^T \textbf{D}$, where $\Lambda=[\lambda_i]_{i=1:N} \in \mathbb{R}^N$ is now a vector of Lagrange multipliers varied by the quality map $\m$ instead of a constant scalar while the vectorized distortion $\textbf{D} \in \mathbb{R}^N$ measures each pixel's distortion.
Each element $\lambda_i$ in $\Lambda$ is determined by a corresponding $m_i$ by a predefined and monotonically increasing function $T:[0,1]\rightarrow\mathbb{R^{+}}$, \ie, $\lambda_{i} = T(m_{i})$. 
In other words, the higher quality level $m_{i}$ is, the higher weight $\lambda_{i}$ is for the distortion term of the corresponding pixel $x_{i}$.
This framework leads to an explicit control over spatial bit allocation guided by $\m$.

The estimation of rate $R$ requires to learn an entropy model $P$ conditioned on $\m$, which outputs the likelihood of the given representation $\hat{\y}$, and replace $R$ by an approximate entropy. 
By choosing squared error as the distortion metric, the training loss of the proposed spatially-adaptive variable-rate image compression model is given by
\begin{equation}
  \label{eqn:pix_weighted_rate_distortion}
  \mathcal{L}=-\log P(\hat{\y}|\m) + \sum_{i}^{N}\lambda_{i}\frac{(x_{i} - {x}'_{i})^{2}}{N}.
\end{equation}
Note that a spatially uniform quality map regularizes the standard rate-distortion optimization with a single scalar $\lambda$.

\begin{figure*}[t]
  \centering
  \includegraphics[width=0.99\linewidth]{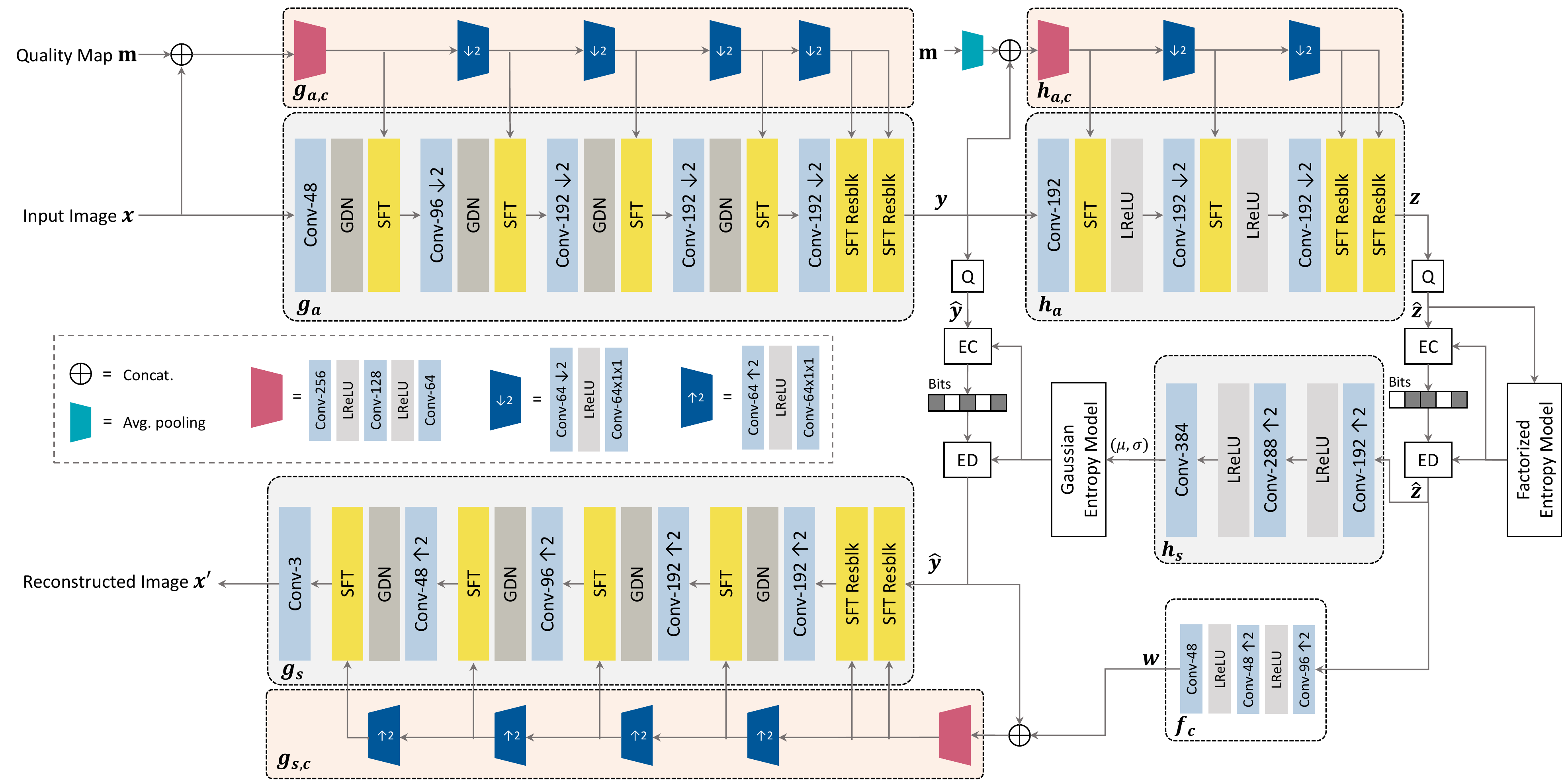}  \\
   \vspace{0.2cm}
  \caption{The network architecture of the proposed model. 
           We insert the SFT layers into the convolutional autoencoder, and utilize several condition networks to generate prior condition features from the input image and the quality map.
           Each of the encoder $g_{a}$, the hyper-encoder $h_{a}$, and the decoder $g_{s}$ has its own conditon network $g_{a,c}$, $h_{a,c}$ and $g_{s,c}$, respectively.
           Note that $h_{s}$ indicates the hyper-decoder.
           We use $3\times3$ kernels for the convolution layers in the condition networks and the SFT modules, and $5\times5$ kernels for others.
           We adopt a simplified version of generalized divisive normalization (GDN) and inverse GDN (IGDN)~\cite{johnston2019computationally} in our network.
           EC and ED denote the entropy coding and entropy decoding, respectively.
          }
    \label{fig:architecture}
\end{figure*}

\subsection{Spatially-adaptive feature transform}
\label{sec:sft_module}
Our network is characterized by Spatial Feature Transform (SFT) module~\cite{wang2018recovering}, for which a condition network generates its appropriate inputs using an external prior. 
We design the specialized SFT modules to image compression, which effectively reflects a quality map for generating desirable compressed representations.
Figure~\ref{fig:sft} illustrates the revised structure of our SFT module.
Note that the SFT module learns to generate a set of element-wise affine parameters $(\gamma, \beta$) for an intermediate feature map $\mathbf{f}$, depending on an external condition $\Psi$; the SFT layer learns a mapping function $\zeta: \Psi\mapsto(\gamma, \beta)$.
Within the layer, a feature transformation is given by
\begin{equation}
  \text{SFT}(\mathbf{f},\Psi) = \gamma \odot \mathbf{f} + \beta,
\end{equation}
where $\odot$ denotes element-wise multiplication.

In the earlier work of SFT~\cite{wang2018recovering}, all SFT layers share a condition network and employ only the external information, \ie, $\m$ in our case, as the external prior.
In our algorithm, the image compression components such as encoder and decoder adopt their own condition networks, which take the inputs of the components in addition to the external prior and generate suitable spatial importance conditions $\Psi$ for the SFT modules.
Moreover, we incorporate the hierarchical SFT layers with progressive downsampling (or upsampling) of the condition features. 
Also, we transform the condition features to a proper size using convolutions instead of simple methods such as average pooling~\cite{kim2020transfer} and nearest-neighbor downsampling~\cite{park2019semantic}. 
These adjustments turns out to improve the capacity of our model and leads to performance gain.

\begin{figure*}[ht]
  \centering
  \begin{tabular}{ccc}
  \hspace{-0.3cm}
    \includegraphics[width=.32\textwidth]{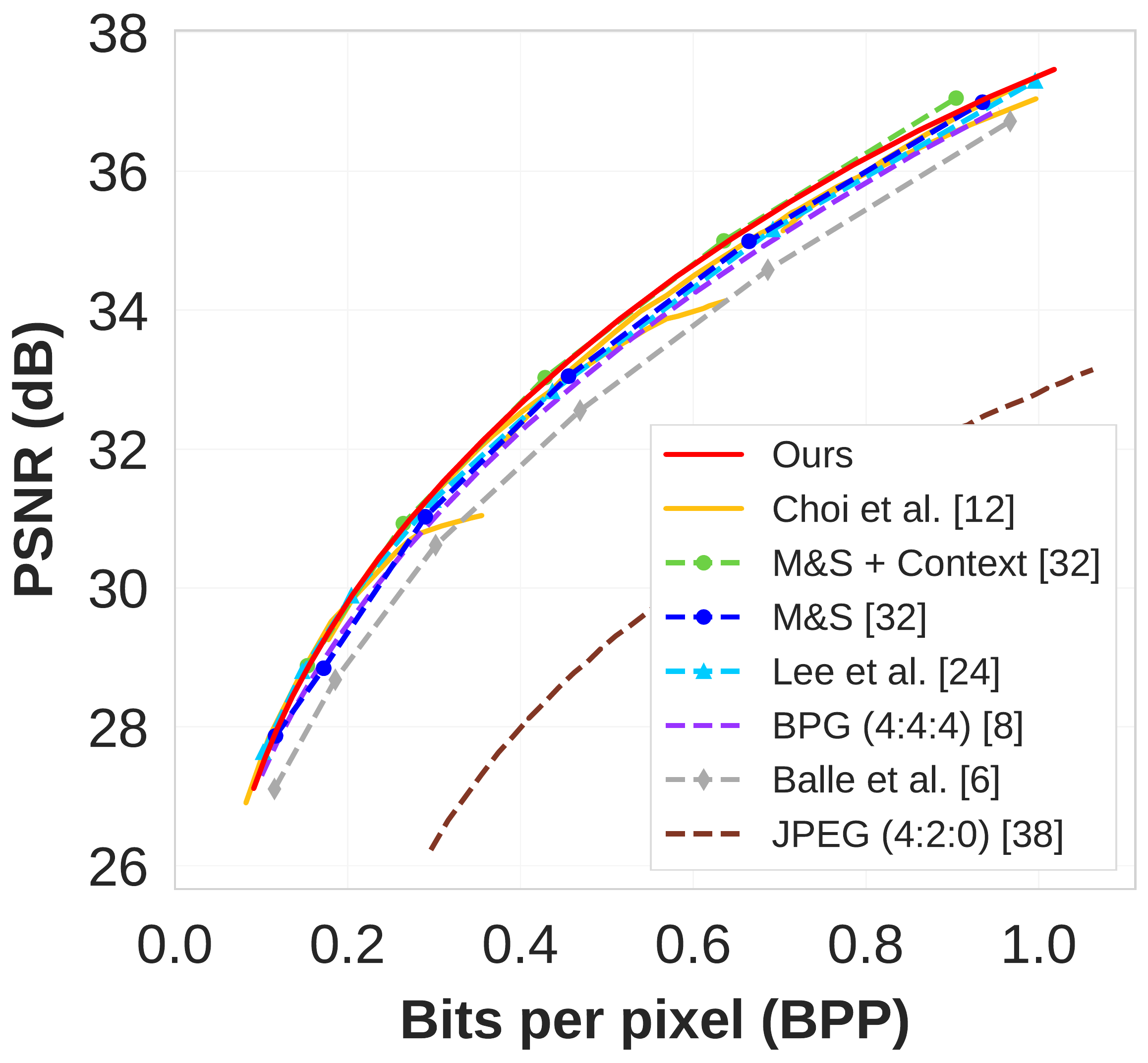} & 
    \includegraphics[width=.32\textwidth]{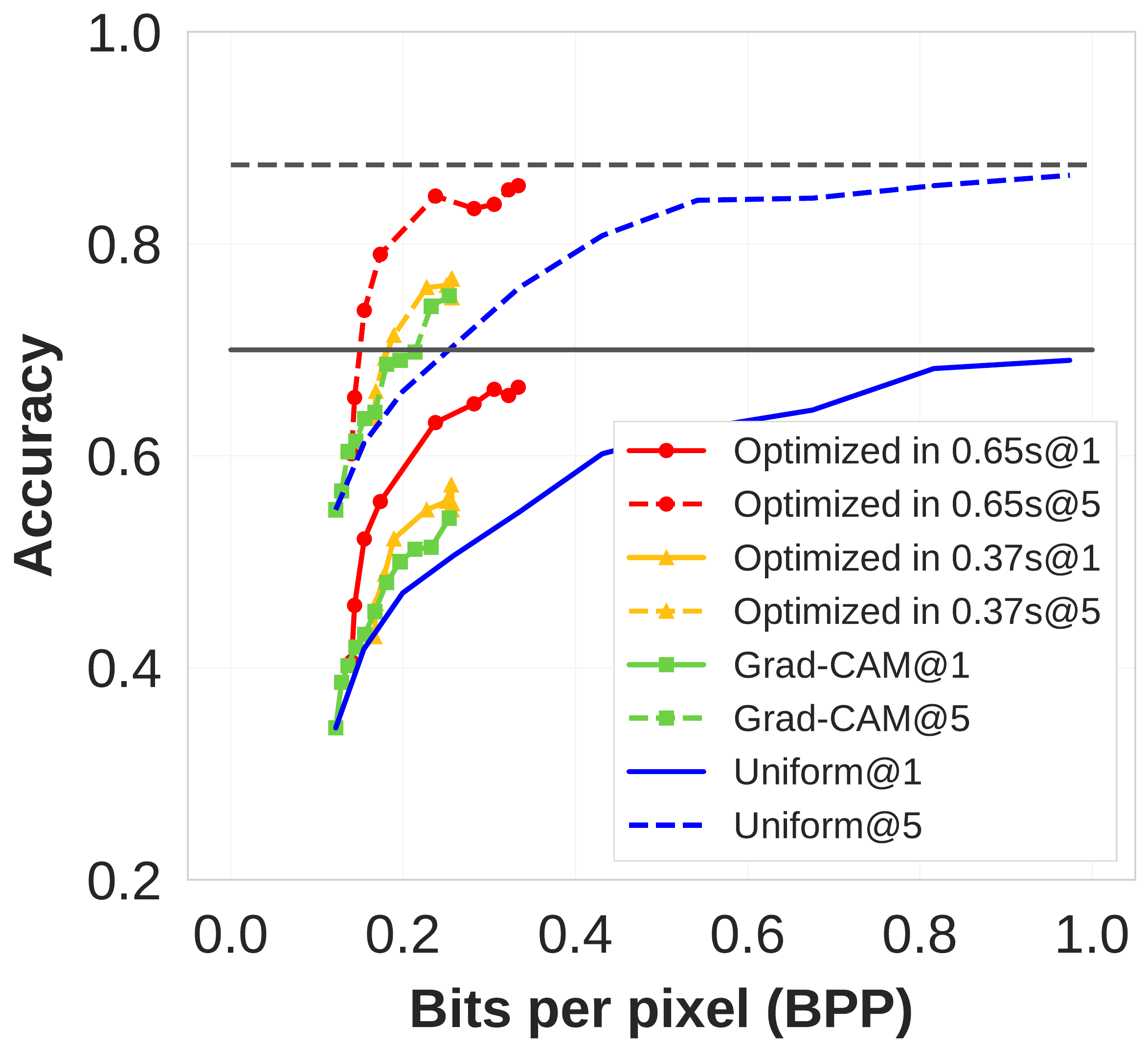} & 
    \includegraphics[width=.32\textwidth]{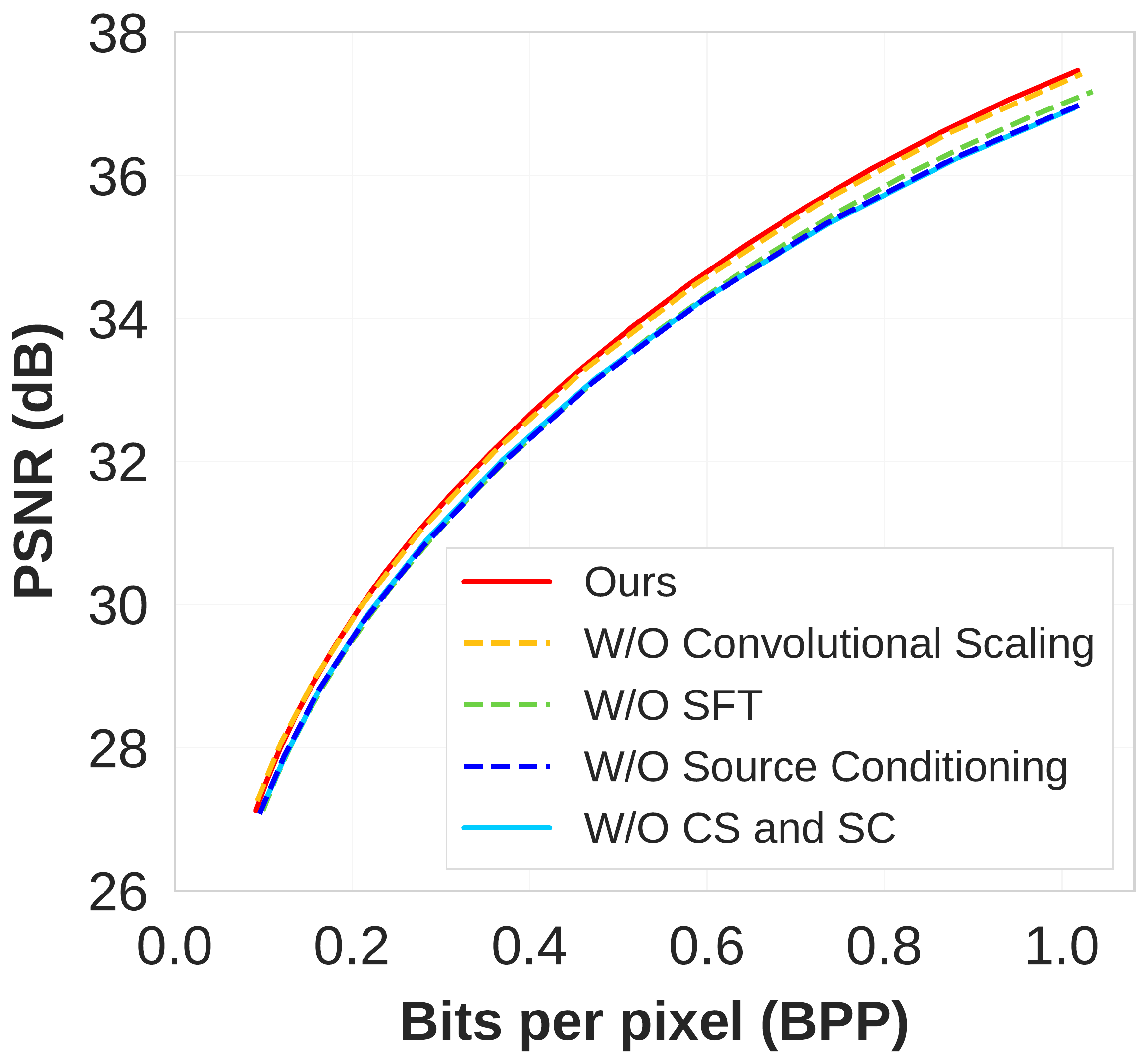}  \\

    (a) & (b) & (c) \\
  \end{tabular}
  \caption{
    (a) PNSR comparison to baseline methods on the Kodak dataset. 
    Mean \& Scale (M\&S) Hyperprior model without context model~\cite{minnen2018joint} is our counterpart single-rate model.
    (b) Classification accuracy comparison on the ImageNet dataset using inferred qualtiy maps and uniform quality maps. 
    The gray lines indicate the accuracies for the original images.
    When we optimize the rate-classification loss three and five times with randomly initialized quality maps, the run times are 0.37 and 0.65 seconds in average, respectively.
    We obtain the rate-accuracy curves for the optimized quality maps by varying the weight of the cross-entropy loss.
    As an alternative of the optimization, we adopt Grad-CAM~\cite{selvaraju2017grad} without the ground-truth label as a quality map.
    (c) Ablation study results.
    W/O Convolutional Scaling (CS) means the replacement of the convolution layers with average pooling for scaling the intermediate features in the condition networks.
    In the case of W/O Source Conditioning (SC), all condition networks take only the external inputs, \ie, $\Psi_1 = g_{a,c}(\m), \Psi_2 = h_{a,c}(\m)$, and $\Psi_3 = g_{s,c}(\mathbf{w})$.
    Lastly, for W/O SFT, we eliminate all the SFT layers and feed the condition features to the preceding convolution layers.
  }
  \label{fig:result_graphs}
\end{figure*}

\subsection{Network architecture and pipeline}
\label{sec:net_architecture}

Figure~\ref{fig:architecture} demonstrates the proposed architecture derived from the Mean~\&~Scale (M\&S) Hyperprior model~\cite{minnen2018joint}.
Among several variations of the models introduced in~\cite{minnen2018joint}, we choose the non-autoregressive version without a context model for our entropy model $P$.
On top of the baseline network, our approach incorporates a condition network with multiple SFT modules for each of the three components---encoder, decoder, and hyper-encoder, where an SFT module is given a $\Psi$ produced by the matching condition network.
We present the detailed procedure of our network designed for image compression in the following.

We first encode an image $\x$ using an encoder $g_a(\cdot, \cdot)$ and a condition network $g_{a,c}(\cdot, \cdot)$ given a quality map $\m$ and generate a latent representation $\y$, which is given by
\begin{align}
  \label{eqn:encoding}
   \y = g_a(\x, \Psi_1),  ~~\text{where}~~  \Psi_1 = g_{a,c}(\x, \m).
\end{align} 
The spatially-adaptive quality information is captured in $\y$.
A hyper-encoder $h_a(\cdot, \cdot)$ generates an image-specific side-information $\z$ from the latent representation $\y$, where another condition network $h_{a,c}(\cdot, \cdot)$ is applied to $(\y, \m)$ and produces a condition variable $\Psi_2$ as follows:
\begin{align}
  \label{eqn:hyper_encoding}
  \z &= h_a(\y, \Psi_2), ~~\text{where}~~ \Psi_2 = h_{a,c}(\y, \m).
\end{align}
Note that $\z$ captures the spatial dependencies in the quantized latent representation $\hat{\y}=Q(\y)$ and models the probability of $\hat{\y}$ conditionally.
Then quantized side information $\hat{\z}=Q(\z)$ is forwarded to the hyper-decoder $h_s(\cdot)$ to draw the parameters $(\mu, \sigma)$ of a Gaussian entropy model, which approximates the distribution of $\hat{\y}$. 

For reconstructing the image, a decoder $g_s(\cdot, \cdot)$ operates on $\hat{\y}$ and the output of the condition network $g_{s,c}(\cdot, \cdot)$, which is given by
\begin{align}
  \label{eqn:decoding}
  \mathbf{x}' = g_s(\hat{\mathbf{y}},\Psi_3), ~~\text{where}~~ \Psi_3 = g_{s,c}(\hat{\y},\mathbf{w}).
\end{align}
Note that a resizing (upsampling) network $f_c(\cdot)$ is applied to $\hat{\z}$ to obtain a surrogate $\w$ of a quality map $\m$ since $\m$ may not be available in the decoder side but $\hat{\z}$ maintains the spatial importance information.

\subsection{Task-aware image compression}
The proposed algorithm provides task-aware image compression capability, which can be done by estimating a task-specific quality map using a pretrained task model at encoding time. 
Given a task loss function $\mathcal{L}_\text{task}$, the optimal task-specific quality map $\m^{*}$ is given by the following objective:
\begin{equation}
  \label{eq:rate_task}
  \m^{*} = \argmin_\m P(\hat{\y}|\m) + \lambda \mathcal{L}_\text{task},
\end{equation}
where $\lambda$ is a Lagrange multiplier for rate-task trade-off control.
Since each element of $\m$ is in $[0, 1]$, the optimization is simply realized by the standard backpropagation.
 
The pretrained task model can be any network for arbitrary tasks even including the third-party ones.
For example, one may acquire a classification-specific quality map to improve the quality of particular semantic regions by choosing a proper pretrained classifier for $\mathcal{L}_\text{task}$ of \eqref{eq:rate_task}.
We emphasize that this additional optimization at test time encourages the candidate quality map to boost the performance in terms of the target task and it is sufficient to use a uniform quality map for decent reconstruction quality in general. 
Also, as a practical estimate of $\m^{*}$, one may use a saliency map from the task model or draw a location-wise information using the intermediate features or outputs of the models, \eg ROI masks.
Nonetheless, our optimized quality maps lead to higher rate-task performance than competing alternatives.
Note that our framework deals with arbitrary tasks by only changing $\m$ at encoding time unlike other task-aware approaches that require task-specific training~\cite{choi2020task}.
The appendix has more discussions.

%% file: sections/Experiments.tex
% !TEX root = ./../submission.tex

%-------------------------------------------------------------------------

\section{Experiments}
\label{sec:experiments}

We now present the experimental results of our compression model in comparison to the existing ones and analyze our model in various conditions.
Refer to the appendix for more discussions and illustrations about our results including training quality map, model complexity, qualitative results, and experimental details.

\begin{figure}[t]
  \centering
  \begin{tabular}{cc}
    \sf{\small ROI Mask} & \sf{\small Uniform} \\
    % {\small ROI Mask} & {\small Uniform} \\
    \includegraphics[width=0.45\linewidth]{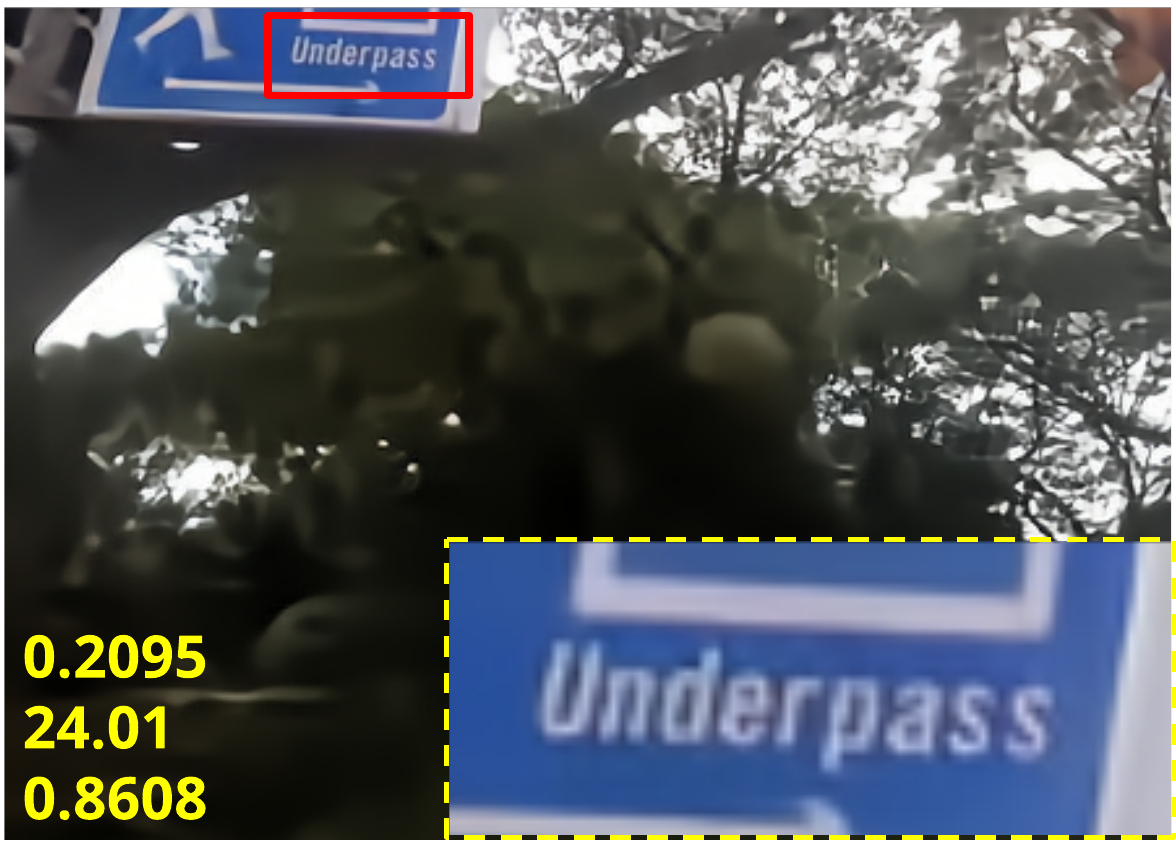} &
    \includegraphics[width=0.45\linewidth]{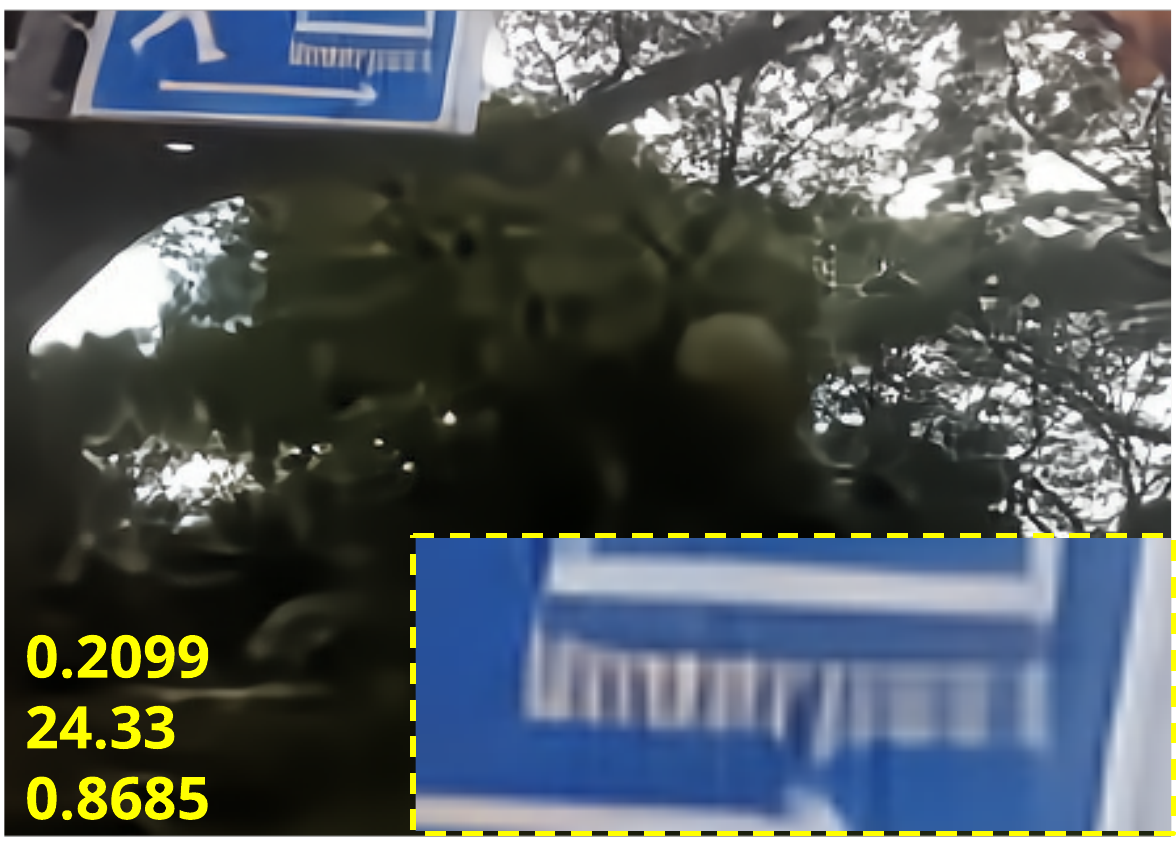} \\
    \includegraphics[width=0.45\linewidth]{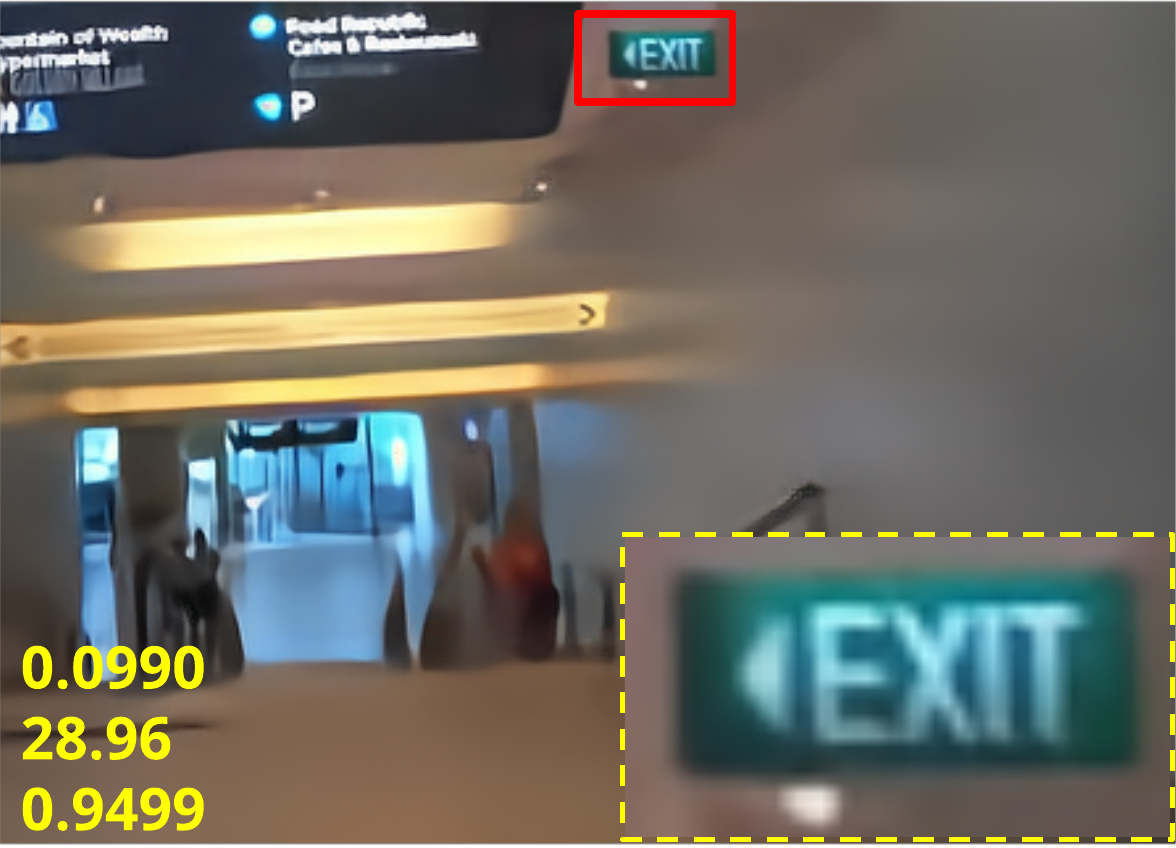} &
    \includegraphics[width=0.45\linewidth]{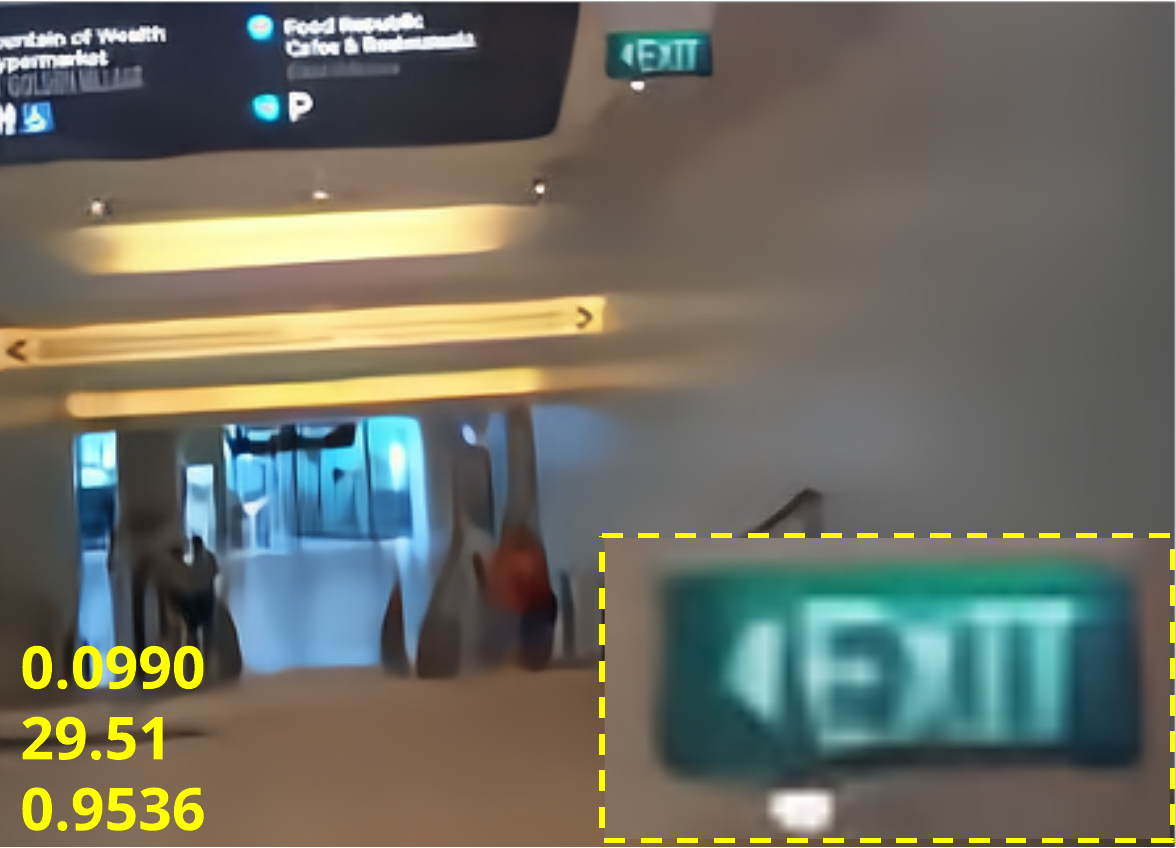} \\
  \end{tabular}
  \vspace{0.1cm}
  \caption{
    Text-preserving compression examples with ROI masks in our model.
    In the left column, ROI mask (red box) highlighting the text region is adopted as a quality map, whereas a uniform quality map is used in the right column.
    The yellow annotations denote the triples of bpp, PSNR (dB), and MS-SSIM.
  }
  \label{fig:qa_text}
\end{figure}

\subsection{Training and evaluation}
\paragraph{Datasets and data processing}
We train our model on the COCO~\cite{lin2014microsoft} dataset with data augmentation via random cropping of $256\times256$ images without resizing.
We evaluate the rate-distortion performance on the Kodak dataset~\cite{kodak}.
For the classification-aware compression, we construct a test set using ImageNet~\cite{russakovsky2015imagenet} by choosing 102 categories and sampling 5 images per category randomly.
We adopt the ICDAR2015~\cite{karatzas2015icdar} dataset to obtain the text-preserving compression results.

\vspace{-0.2cm}
\paragraph{Implementation details}
Our implementation relies on Pytorch~\cite{paszke2019pytorch} and an open-source compression library~\cite{begaint2020compressai}.
For the entropy coder, we adopt Range Asymmetric Numeral System~\cite{duda2013asymmetric} provided by \cite{begaint2020compressai}.
We train our model is trained for 2M iteratrions with batch size 8.
We employ the Adam~\cite{kingma2015adam} optimizer, where the learning rate is initially set to be $10^{-4}$ and decreased to $10^{-5}$ after 1.4M iterations.
Gradient clipping~\cite{pascanu2013difficulty} with threshold 1 leads to stable training according to our experience.
The function $T(\cdot)$, which produces $\Lambda$ from the quality map, is defined as $T(x)=0.001e^{4.382x}$; this choice approximately results in the bpp range of $[0.1, 1.0]$ on the Kodak dataset.

\subsection{Rate-distortion performance}
We first evaluate the performance of our variable-rate model by feeding multiple uniform quality maps with no consideration of spatial adaptivity.
For comparisons, we select recent learning-based image compression models~\cite{balle2018variational,minnen2018joint,lee2018context,choi2019variable} and the classical state-of-the-art codec, BPG~\cite{bellard2014bpg}.
PSNR is employed as an evaluation metric since all these models are optimized for MSE.
This experiment is conducted on the Kodak dataset.

Figure~\ref{fig:result_graphs}(a) illustrates the superior quality of our model compared to the baseline methods. 
Unlike other variable-rate techniques~\cite{choi2019variable,yang2020variable}, which achieve slightly worse or similar performances compared to the single-rate counterparts under the equivalent architectures of the entropy model, our model outperforms the corresponding single-rate model, M\&S~\cite{minnen2018joint}.
Our model is even better or as competitive as Lee~\etal~\cite{lee2018context} and M\&S+Context~\cite{minnen2018joint}, which adopt the time-consuming autoregressive context models.
Note that the single-rate approaches require training multiple independent models to cover a wide range of rates, \eg, 6 models of M\&S in Figure~\ref{fig:result_graphs}(a).
Our algorithm also outperforms the recent variable-rate method~\cite{choi2019variable} with the autoregressive context model.
We emphasize that our model does not include the context model and has potential to further improve performance with it.
%Additional results including model complexity are provided in the appendix.

\subsection{Task-aware compression}

\paragraph{Classification} 
Figure~\ref{fig:result_graphs}(b) compares classification accuracy with three options of quality maps including inferred maps by rate-task optimization, Grad-CAM~\cite{selvaraju2017grad}, and uniform quality maps.
We employ the pretrained VGG16~\cite{simonyan2015very} to optimize the cross-entropy loss for the rate-task optimization and extract the Grad-CAM outputs.
The class with the highest prediction score in each image is selected to estimate Grad-CAM instead of the ground-truth label since the scenario is more practical.
To validate the generalization performance, we use pretrained ResNet18~\cite{he2016deep} to obtain classification accuracy.
Our task-aware quality map optimization at test time significantly improves classification accuracy while incurring less than a second computational cost with a TitanXp GPU.
Grad-CAM also outperforms the uniform quality maps, which demonstrates the feasibility of the task-aware compression without the ground-truth labels.

\vspace{-0.2cm}
\paragraph{Text preserving}
Figure~\ref{fig:qa_text} demonstrates text-preserving compression results with manually given text ROI masks.
The results imply that our model successfully reflects the ROI information and preserves the text regions even at harsh compression rates. 

\subsection{Qualitative analysis}

\begin{figure*}[t]
  \centering
  \includegraphics[width=\textwidth]{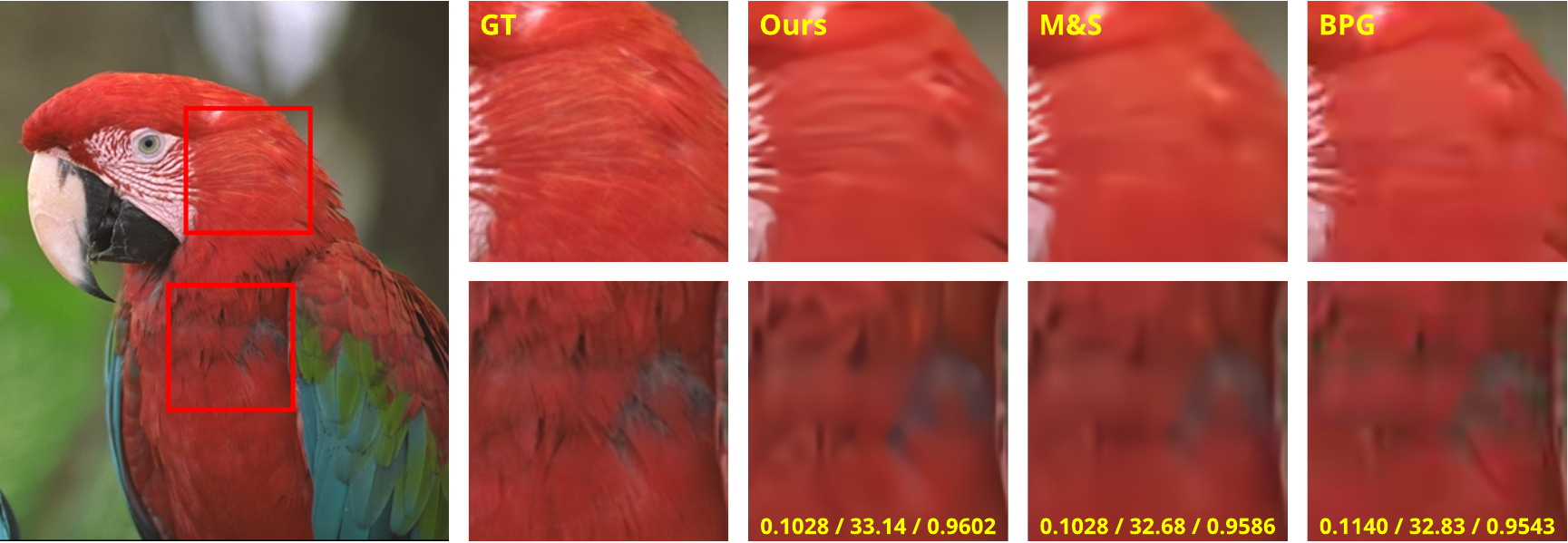}
  %\vspace{0.2cm}
  \caption{Comparison between our model, M\&S Hyperprior model~\cite{minnen2018joint}, and BPG~\cite{bellard2014bpg}.
  The yellow annotations in the cropped images indicate bpp/PSNR (dB)/MS-SSIM of an entire image.}
  \label{fig:qa_comparison}
\end{figure*}

\begin{figure*}[t]
  \centering
  \includegraphics[width=\textwidth]{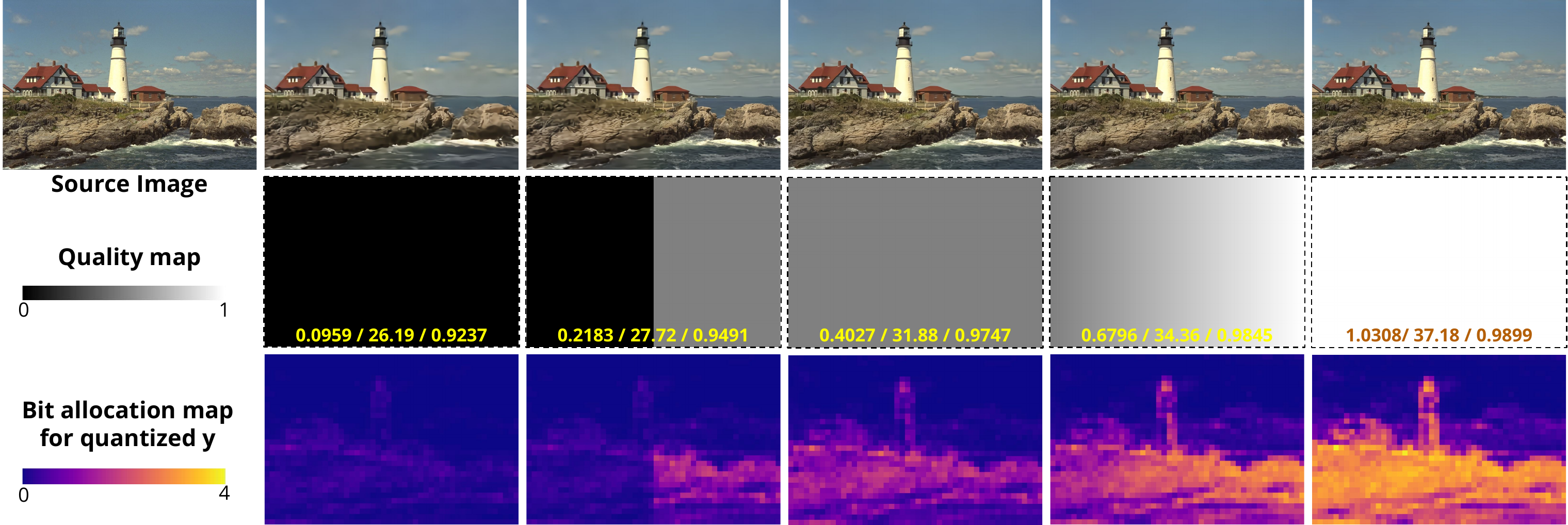}
  %\vspace{0.2cm}
  \caption{Our spatially-adaptive compression results for uniform and non-uniform quality maps with the corresponding bit allocation maps.
           The yellow numbers in the quality map denote bpp/PSNR (dB)/MS-SSIM.
           The reconstructed images are best views with zooming-in.
           }
  \label{fig:qa_bitsmap}
\end{figure*}

\paragraph{Comparison with other methods}
Figure~\ref{fig:qa_comparison} shows our result with a uniform quality map in comparison with M\&S Hyperprior model~\cite{minnen2018joint} and BPG~\cite{bellard2014bpg} on an image from the Kodak dataset.
Our method reconstructs complex textures better than others while achieving the best PSNR and MS-SSIM scores at lower or the same bitrate. 

\vspace{-0.2cm}
\paragraph{Quality map vs. bit allocation}
Figure~\ref{fig:qa_bitsmap} illustrates the reconstruction and bit allocation results for different quality maps on an image from the Kodak dataset.
The bit allocation maps are given by the average negative log-likelihood of $\hat{\y}$ over all channels at every element.

\subsection{Ablation Study}
\label{sec:ablation}

Figure~\ref{fig:result_graphs}(c) presents the results from the ablation study to validate the contribution of each module on rate-distortion performance; the proposed approach adopting the SFT layers and refining the modules improves the quality of the variable-rate image compression.
Most importantly, comparing our model with W/O Source Conditioning case, additional conditioning on the original input of the network as well as the external data, \eg, $\x$ and $\m$ for $g_{a,c}$, contributes to the performance noticeably.
This is because the adaptivity of our model is enhanced by the generation of different condition features depending on the original image given the same quality map.

%% file: sections/Conclusion.tex
% !TEX root = ./../submission.tex

%-------------------------------------------------------------------------
\section{Conclusion}
\label{sec:conclusion}

We presented a novel variable-rate image compression algorithm based on an end-to-end trainable deep neural network.
The proposed framework enables us to perform spatially-adaptive image compression in terms of compression rate using a single model given a real-valued pixel-wise quality map.
In addition to the flexibility, our model has an additional benefit to estimate task-specific quality maps automatically at test time without additional training, which is useful to perform task-aware image compression.
We design an efficient network architecture based on spatially-adaptive feature transform for the conditioned image compression.
Our experiment shows that the proposed algorithm achieves outstanding performance even compared to multiple models trained independently with fixed rates and has wide applicability to various tasks in computer vision.

\vspace{-0.2cm}
\paragraph{Acknowledgements} 
This work was partly supported by Samsung Advanced Institute of Technology and the Bio \& Medical Technology Development Program of the National Research Foundation (NRF) funded by the Korean government (MSIT) (No. 2021M3A9E4080782).

%% file: sections/Appendix.tex
% !TEX root = ./submission.tex

\appendix
\section*{Appendix}
\label{sec:appendix}

\section{Additional qualitative results}
\label{sec:appendix_qualitatives}
We present more qualitative results in the following three categories.

\paragraph{Text-preserving}
Figure~\ref{fig:appendix_text} shows examples of text-preserving compression using a ROI mask.

\begin{figure*}[t]
  \centering
  \includegraphics[width=0.75\textwidth]{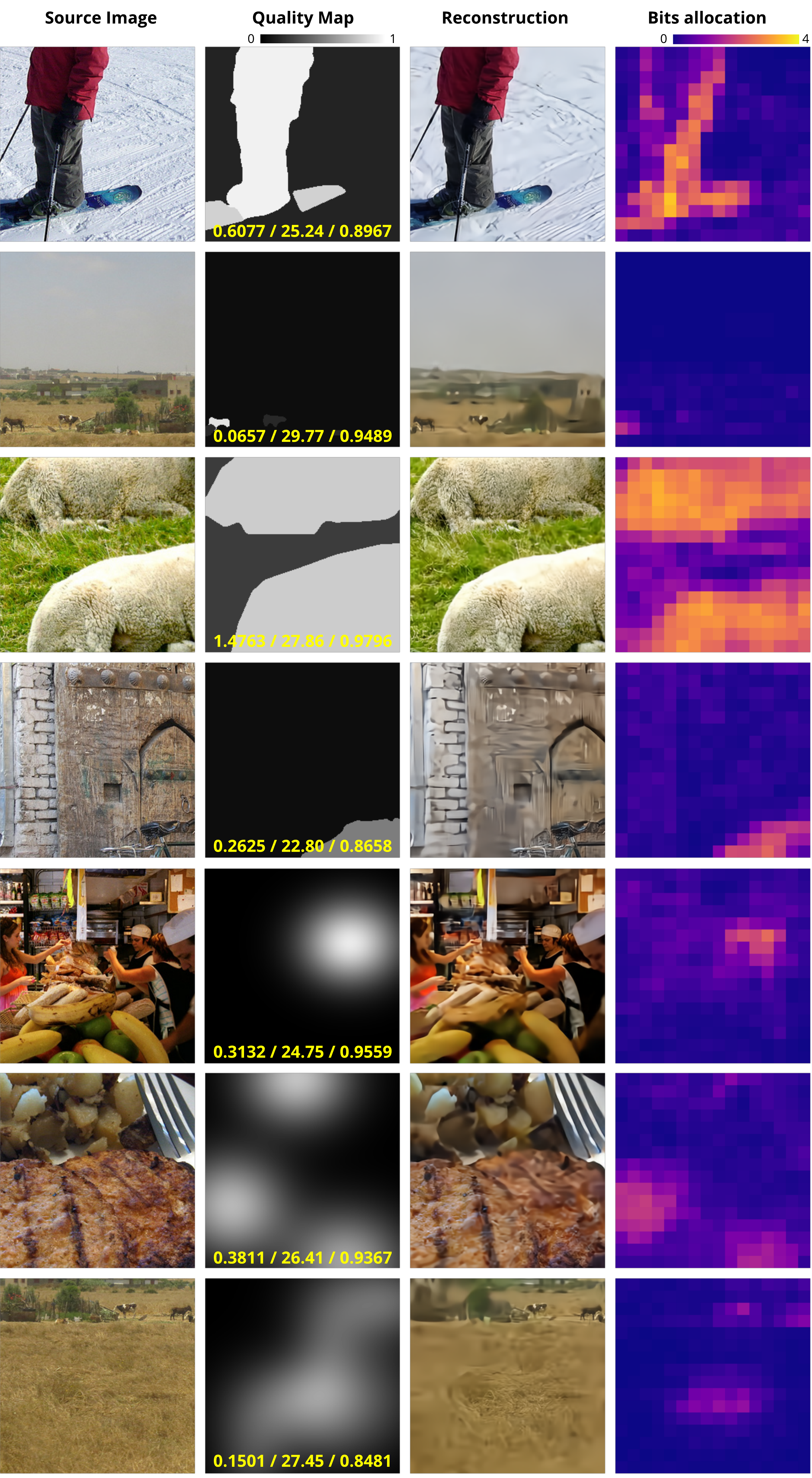}
  \caption{Examples of compression results using non-uniform quality maps.
           }
  \label{fig:appendix_nonuniform}
\end{figure*}

\paragraph{Non-uniform quality maps}
Figure~\ref{fig:appendix_nonuniform} presents compression results using non-uniform quality maps on the COCO~\cite{lin2014microsoft} validation set and the average bit allocation maps of quantized latent representation $\hat{y}$.
All these example images were not included in our training set since we trained our model using the COCO train split.

\paragraph{Uniform quality maps}
Figure~\ref{fig:appendix_kodak_5}, \ref{fig:appendix_kodak_14}, \ref{fig:appendix_kodak_19}, \ref{fig:appendix_kodak_22} show qualitative comparison results of our model, Mean \& Scale (M\&S) Hyperprior model~\cite{minnen2018joint}, BPG (4:4:4)~\cite{bellard2014bpg} and JPEG (4:2:0)~\cite{pennebaker1992jpeg} on several Kodak images.
We adapt the compression rate of our model to that of M\&S Hyperprior model by adjusting the value of the uniform quality map.
Our model outperforms all other methods in terms of the visual quality and PSNR/MS-SSIM scores at similar bit rates.

% ------------------------------------------------------------------------------

\section{Additional RD performance comparisons}
\label{sec:appendix_rd}

\begin{figure}[t]
  \centering
  \includegraphics[width=0.8\linewidth]{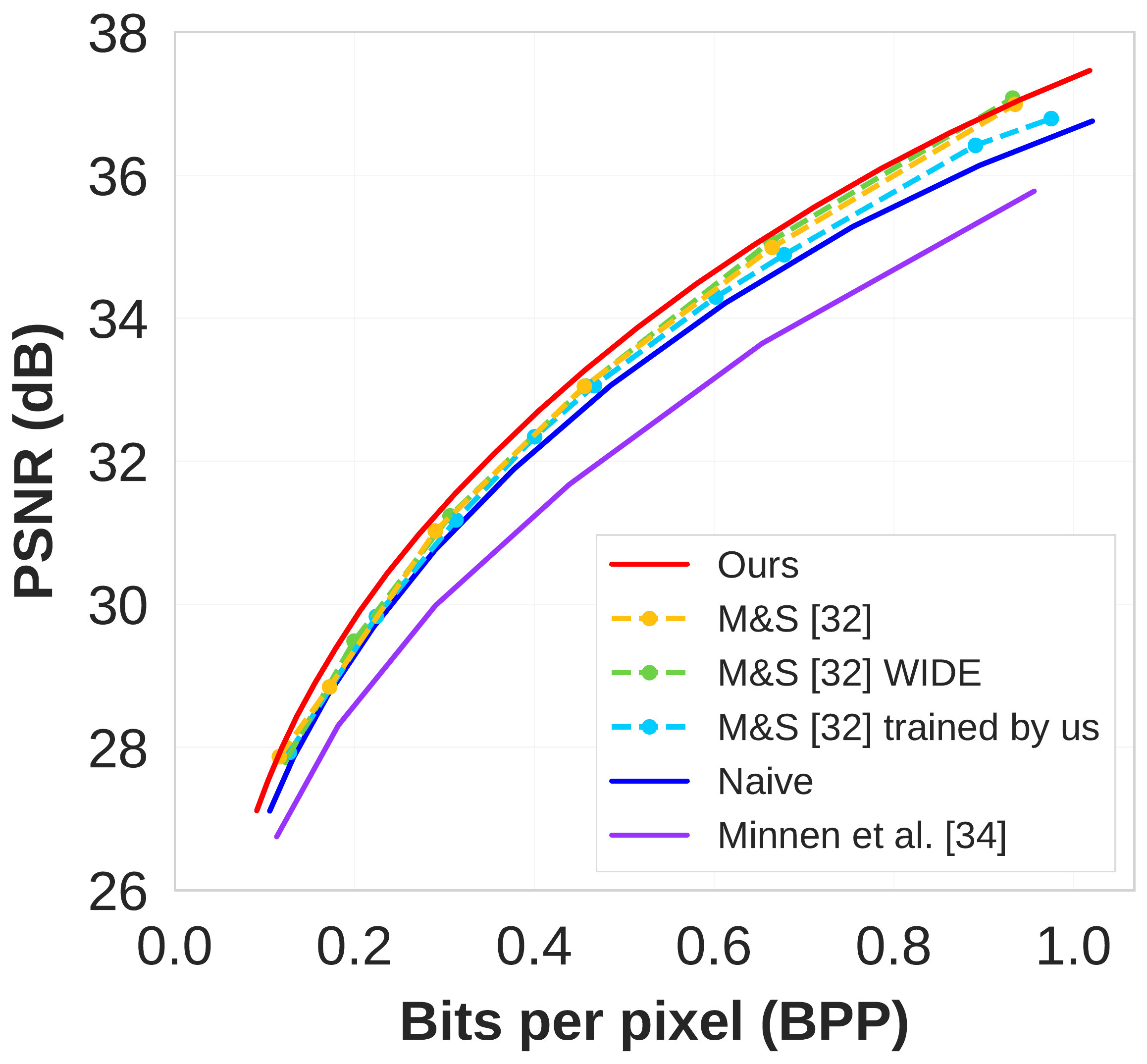}
  \caption{
    The rate-distortion performances on Kodak dataset.
    ``M\&S WIDE'' is a wide-layer version of M\&S Hyperprior model which has the similar number of parameters to that of our model.
    ``Naive'' is a naive implementation of spatially-adpative variable-rate model which is based on M\&S Hyperprior model.
  }
  \label{fig:appendix_rd}
\end{figure}

To show the proposed model is particularly suitable for our task, we implement a variant of M\&S Hyperprior model~\cite{minnen2018joint} as a naive spatially-adpative variable-rate approach and present it in Figure~\ref{fig:appendix_rd}.
The naive model has same architecture as M\&S Hyperprior model, but $(\x, \m)$, $(\y, \m)$ and $(\hat{\y}, \w)$ are inputs to the encoder, hyper-encoder and decoder of it, respectively.
We trained it with the same training setting as our model was trained with.
We also report the official performances of M\&S Hyperprior model, Minnen~\etal~\cite{minnen2017spatially} which allows per-patch quality adaptation, and the performance of M\&S Hyperprior model trained by us for 2M iterations.

Compared to M\&S Hyperprior model trained by us, the naive model shows performance degradation, which implies that the naive approach is not sufficient for the task of the image compression with quality map.
Minnen~\etal~\cite{minnen2017spatially} shows poor quality compared to other models.
Meanwhile, we observe a slight drop in the performance of M\&S Hyperprior model trained by us in comparison with the officially reported performance of M\&S Hyperprior model.
It may be because the original model was trained for about 6M iterations as the authors mentioned\footnote{\url{https://groups.google.com/g/tensorflow-compression/c/LQtTAo6l26U/m/cD4ZzmJUAgAJ}}.

% ------------------------------------------------------------------------------

\section{Model complexity}
\label{sec:appendix_complexity}

\begin{table}[t]
  \centering
  \caption{Comparison of model complexity on Kodak dataset for our model and the baseline models using a GPU. 
  }
  \vspace{0.03cm}
  \scalebox{0.7}{
  \begin{tabular}{c|cccc}
    & \# Parameters & Rate & Encoding (ms) & Decoding (ms) \\ \hline
    M\&S~\cite{minnen2018joint}      & 11M & single & 55  & 31    \\
    M\&S~\cite{minnen2018joint} WIDE & 28M & single & 143 & 45    \\
    M\&S + Context~\cite{minnen2018joint} & 14M & single & 5004 (82) & 9876    \\
    Choi~\etal~\cite{choi2019variable} & 37M & variable & 8004 (185) & 12973       \\
    Ours W/O SC   & 27M & variable & 250 & 37         \\
    Ours            & 28M & variable & 254 & 36     \\
    
  \end{tabular}}
  \vspace{-0.45cm}
  \label{table:complexity}
\end{table}

\begin{figure*}[t]
  \centering
  \includegraphics[width=0.8\linewidth]{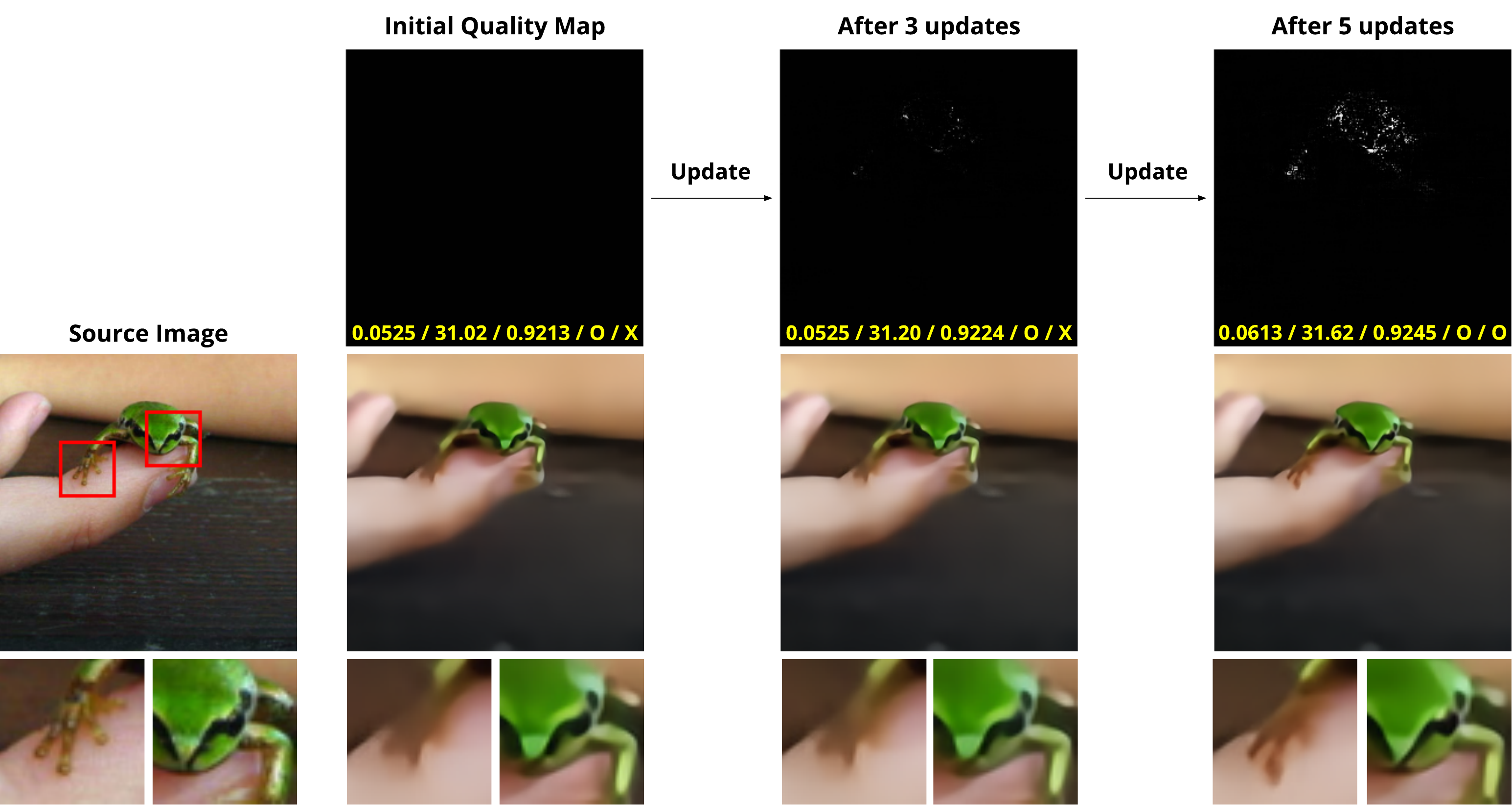}
  \caption{
    Classification-aware quality maps and recontructions through iterations when $\lambda = 0.01$.
    The annotation in each quality map denotes bpp/PSNR (dB)/MS-SSIM/Top-5/Top-1 classification result of the corresponding reconstructed image.
  }
  \label{fig:appnedix_classification_update}
\end{figure*}

Table~\ref{table:complexity} compares the number of parameters and average encoding/decoding runtimes of our model and the baseline models.
We additionally trained a wide-layer version of M\&S Hyperprior model (304 channels) and denote it as M\&S WIDE.
We used a machine with a Titan Xp GPU and all models utilized the same entropy coder.

The M\&S family requires multiple independent models to cover wide range of rate (\ie, 6 models of M\&S in Figure~\ref{fig:appendix_rd}), thus the actual number of the parameters is the multitude of the numbers in Table~\ref{table:complexity}.
M\&S + Context~\cite{minnen2018joint} and Choi~\etal~\cite{choi2019variable} require much coding time than our model since they employ the autoregressive context model which have a serial coding process.
The encoding of such serial context models can be made efficient by using masked convoltuion, \eg, for M\&S + Context the encoding time decreased from 5004 ms to 82 ms.
However, the decoding is inevitably slow and cannot utilize parallel processing.
Our model outperforms M\&S WIDE which has the similar number of parameters to ours as in Figure~\ref{fig:appendix_rd}.
The coding complexity of our approach increases marginally compared to the version without source conditioning (W/O SC) while the performance improves significantly.
These results imply that simply increased parameters do not necessarily lead to performance gain while the proposed SFT with SC effectively improves the capacity of compression model.

% ------------------------------------------------------------------------------

\section{Experiment details}
\label{sec:appendix_experiment_details}

\subsection{Rate-distortion comparison}
This section describes how we obtain the plots in Figure~\ref{fig:result_graphs}(a) in detail.
For our model, we used a set of $q$-valued uniform quality maps ($q\in\{0, 0.05, 0.10, ..., 1.0\}$) and computed the average metrics over test images for each of 21 quality maps.
Smaller spacing of $q$ led to almost identical curves.
For Choi \etal~\cite{choi2019variable}, we extracted the RD curves from the original paper.
For other methods, we used the results\footnote{\url{https://github.com/tensorflow/compression/tree/master/results/image_compression}} provided by authors of \cite{balle2018variational,minnen2018joint}.

\subsection{Classification-aware compression}
This section describes how we obtain the plots in Figure~\ref{fig:result_graphs}(b) in detail.
We constructed a test set based on the ImageNet~\cite{russakovsky2015imagenet} dataset by sampling 102 categories and choosing 5 images per a category randomly.
We iteratively updated randomly initialized $\m$ by minimizing the loss for the test set:
\begin{equation}
  \label{eq:rate_classification}
  \mathcal{L}=-\log P(\hat{\y}|\m) + \lambda \mathcal{L}_\text{CE}, 
\end{equation}
where $\mathcal{L}_\text{CE}$ denotes the cross-entropy loss.
We took $\lambda \in \left \{ 0.0001, 0.001, 0.004, 0.01, 0.1, 1, 10, 100, 1000 \right \}$ and used the results at 3 and 5 iterations for each plot.
The accuracies converged to the known upper bound, the accuracies on the original images.
We adopted the L-BFGS~\cite{liu1989limited} solver as an optimizer. 
During optimization, we used a pretrained VGG16~\cite{simonyan2015very} to compute $\mathcal{L}_\text{CE}$ loss while ResNet18~\cite{he2016deep} was used at test time to validate the generalization performance. 
For the Grad-CAM~\cite{selvaraju2017grad} plots, we choose $\m = \alpha \mathbf{CAM}$ as the quality maps with $\alpha \in \left \{ 0, 0.1, 0.2, ..., 1.0 \right \}$, where $\mathbf{CAM}$ denotes the map acquired by Grad-CAM.
Figure~\ref{fig:appnedix_classification_update} visualizes classification-aware image compression results together with how the classification-aware quality maps are acquired through iterations. 

% ------------------------------------------------------------------------------  

\section{Human evaluation}
\label{sec:appendix_human_evaluation}

\begin{table}[t]
  \centering
  \caption{Human evaluation results for 33 people when using semantic ROI masks as quality maps.
  Average response rates for 16 test cases are presented.
  Each number in parentheses indicate the quality value of ROI/non-ROI.}
  %\vspace{0.1cm}
  \scalebox{0.78}{
  \begin{tabular}{cccc}
                & Uniform (0.25/0.25)  & ROI1 (0.65/0.15)   & ROI2 (0.8↑/0.02)  \\ \hline
    Best        & 16.1\%               & 30.5\%             & 53.4\%                \\   
    Worst       & 72.7\%               & 7.6\%              & 19.7\%                \\ 
  \end{tabular}}
  \vspace{-0.6cm}
  \label{table:human_eval}
\end{table}
We conducted a human evaluation to verify that different quality specifications in semantic and background regions can lead to perceptual improvement at same bitrates.
We constructed 16 test cases from MSRA10K~\cite{ChengPAMI} by randomly sampling original images and corresponding ROI masks.
Each test case consists of the orinal image and three reconstructed images compressed with different qualtiy maps which give almost same bitrates.
For each case, we asked 33 people to select the best and worst reconstructed images given the original image.
The average response rates are presented in Table~\ref{table:human_eval}.

We observe that the higher the quality value is used in the semantic region, the more preferences occur as the best perceptual quality at the same bitrate.
Similarly, for the selecitons of the worst images, the ROI-based quality maps lead to better perceptual quality than uniform quality maps.
However, the votes for ROI2 as the worst images are more than those of ROI1, though the quality value in the semantic region of ROI2 is higher than that of ROI1.
It implies that very poor qualtiy of the non-semantic region is not negligible for human perception.

% ------------------------------------------------------------------------------  

\begin{figure}[t]
  \centering
  \includegraphics[width=1.0\linewidth]{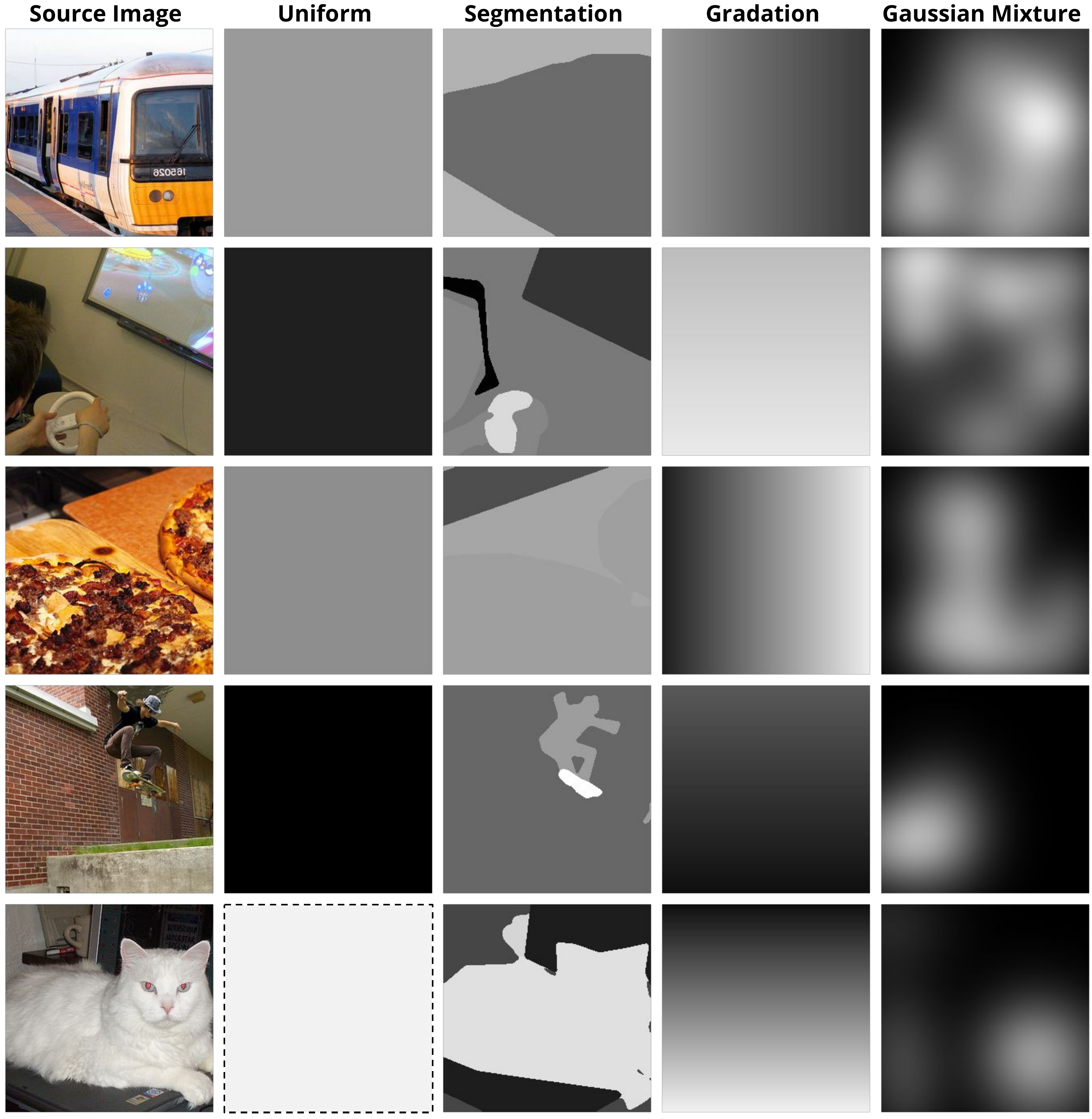}
  \caption{Examples of quality maps used for training.
  For each instance in a mini-batch, we randomly generated a quality map among the four types.}
  \label{fig:training_qmap}
\end{figure}

\section{Example quality maps for training}
\label{sec:appendix_training_qmap}

Figure \ref{fig:training_qmap} presents examples of quality maps we used for training.
Specifically, we randomly generate the quality maps using one of four different ways for each instance in a mini-batch; 
(1) a uniform map
(2) a semantic map of which each class label is converted to random value
(3) a gradation image between two randomly selected values
(4) a kernel density estimation map of Gaussian mixture with random mean, variance and number of mixtures.

% ------------------------------------------------------------------------------

\section{Practical aspects of task-aware compression}
\label{sec:appendix_task_aware}
\begin{figure*}[t]
  \centering
  \includegraphics[width=0.9\linewidth]{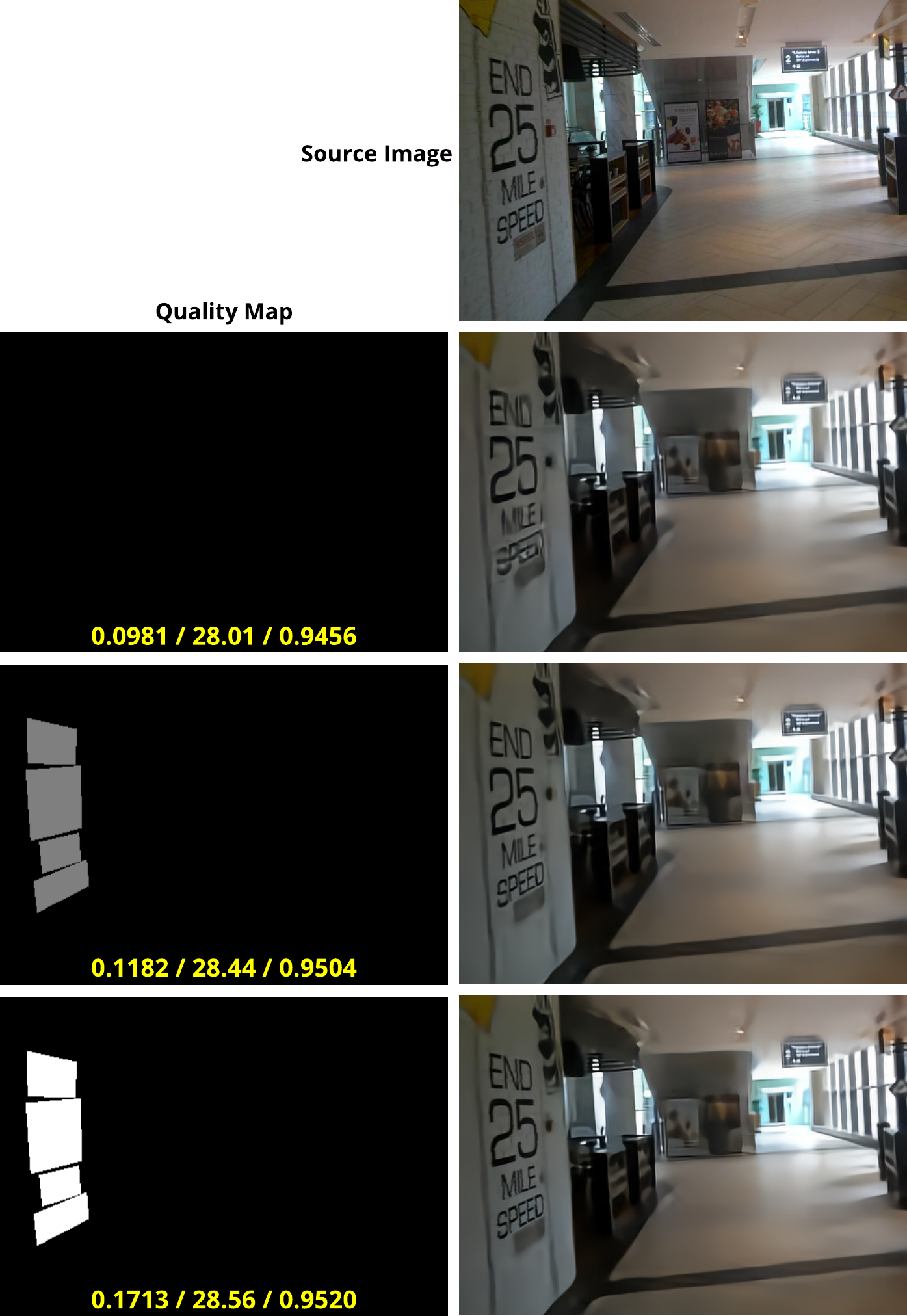}
  \caption{
    Reconstruction results using different quality maps (1\textsuperscript{st} column) for the source image (1\textsuperscript{st} row).
    The annotation in each quality map denotes bpp/PSNR (dB)/MS-SSIM of the corresponding reconstructed image.
    The ROI mask for the text was used as the quality map in 3\textsuperscript{rd} and 4\textsuperscript{th} row.
  }
  \label{fig:appendix_text}
\end{figure*}

One can raise some questions about feasibility and necessity of the task-aware compression; 
How would the task-aware compression be applied in a real-world application when a target label is not available?
Why is the task-aware optimization not redundant when we already have the label? 
Wouldn’t it be just cheaper to store the task outputs instead of optimization for the task?
The goal of the task-aware compression is to reflect one's preference of spatial quality to a compressed image depending on target tasks.
For example, when constructing street view images, one may want to decrease the quality of human faces while improving that of signs. 
In video conference applications, the quality enhancement only in human region may be required.
In this respect, when the quality of particular semantic regions is important, the classification-aware compression would be useful as shown in the 5\textsuperscript{th} column of Figure~\ref{fig:test_image}, \ref{fig:appnedix_classification_update}.
Even if obtaining a ground-truth label or calculating a task loss is unavailable, we can make an appropriate quality map using external task models, \eg, ROI detection result, or Grad-CAM as shown in Figure~\ref{fig:result_graphs}(b), \ref{fig:qa_text}, \ref{fig:appendix_text} instead of optimizing the quality map.
Note that for Grad-CAM in Figure~\ref{fig:result_graphs}(b), we used a class with the highest score predicted by the classifier for each test image instead of the ground-truth.
Thus, the task-aware compression is still practically feasible without the task labels.

Meanwhile, the task-aware compression can be utilized when the task label itself is not a primary concern.
For the example of video conference applications, we do not want to deliver the position of the human in the image or his/her personal information, but want to deliver the compressed images with high quality of the human region at the expense of the quality of background.
Similarly, if we want to preserve an object region well, the classification-aware compression can be used regardless of the target label.

% ------------------------------------------------------------------------------

\newcommand{\qkodak}[8]{
  \begin{figure*}[t]
    \centering
    \begin{tabular}{cc}
       Source Image & \\
       \includegraphics[width=#7\textwidth]{fig/appendix/qa#2/kodak#1_in_box.png} & \\
    
       Ours #3 & M\&S #4 \\
       \includegraphics[width=#7\textwidth]{fig/appendix/qa#2/kodak#1.png_recon.png} & 
       \includegraphics[width=#7\textwidth]{fig/appendix/qa#2/kodak#1_ms.png} \\
       
       BPG #5 & JPEG #6 \\
       \includegraphics[width=#7\textwidth]{fig/appendix/qa#2/kodak#1_bpg.png} & 
       \includegraphics[width=#7\textwidth]{fig/appendix/qa#2/kodak#1_jpeg.jpg} \\
     \end{tabular}
   
     \begin{tabular}{ccccc}
         Ground Truth & Ours & M\&S & BPG & JPEG \\
         \includegraphics[width=#8\textwidth]{fig/appendix/qa#2/kodak#1.png_input_crop_0.png} & 
         \includegraphics[width=#8\textwidth]{fig/appendix/qa#2/kodak#1.png_recon_crop_0.png} & 
         \includegraphics[width=#8\textwidth]{fig/appendix/qa#2/kodak#1_ms_crop_0.png} & 
         \includegraphics[width=#8\textwidth]{fig/appendix/qa#2/kodak#1_bpg_crop_0.png} & 
         \includegraphics[width=#8\textwidth]{fig/appendix/qa#2/kodak#1_jpeg_crop_0.png} \\
   
         \includegraphics[width=#8\textwidth]{fig/appendix/qa#2/kodak#1.png_input_crop_1.png} & 
         \includegraphics[width=#8\textwidth]{fig/appendix/qa#2/kodak#1.png_recon_crop_1.png} & 
         \includegraphics[width=#8\textwidth]{fig/appendix/qa#2/kodak#1_ms_crop_1.png} & 
         \includegraphics[width=#8\textwidth]{fig/appendix/qa#2/kodak#1_bpg_crop_1.png} & 
         \includegraphics[width=#8\textwidth]{fig/appendix/qa#2/kodak#1_jpeg_crop_1.png} \\
       \end{tabular}
    \caption{Compression results including the source image, our result, M\&S Hyperprior model~\cite{minnen2018joint}, BPG (4:4:4) and JPEG (4:2:0) on Kodak #1 image.
            Each number in parentheses indicates bpp/PSNR (dB)/MS-SSIM of the reconstructed image.
            We adapt the compression rate of our model to that of M\&S Hyperprior model by adjusting the value of the uniform quality map.
            Our model outperforms all other methods in terms of the visual quality and PSNR/MS-SSIM metrics at similar bit rates.
            }
    \label{fig:appendix_kodak_#1}
   \end{figure*}
}

\newcommand{\qkodakv}[8]{
  \begin{figure*}[t]
    \centering
    \begin{tabular}{ccc}
     % \hspace{-0.3cm}
       
       Source Image & Ours #3 & M\&S #4 \\
       \includegraphics[width=#7\textwidth]{fig/appendix/qa#2/kodak#1_in_box.png} &
       \includegraphics[width=#7\textwidth]{fig/appendix/qa#2/kodak#1.png_recon.png} & 
       \includegraphics[width=#7\textwidth]{fig/appendix/qa#2/kodak#1_ms.png} \\
       
       & BPG #5 & JPEG #6 \\
       &
       \includegraphics[width=#7\textwidth]{fig/appendix/qa#2/kodak#1_bpg.png} & 
       \includegraphics[width=#7\textwidth]{fig/appendix/qa#2/kodak#1_jpeg.jpg} \\
     \end{tabular}
   
     \begin{tabular}{ccccc}
       % \hspace{-0.3cm}
         Ground Truth & Ours & M\&S & BPG & JPEG \\
         \includegraphics[width=#8\textwidth]{fig/appendix/qa#2/kodak#1.png_input_crop_0.png} & 
         \includegraphics[width=#8\textwidth]{fig/appendix/qa#2/kodak#1.png_recon_crop_0.png} & 
         \includegraphics[width=#8\textwidth]{fig/appendix/qa#2/kodak#1_ms_crop_0.png} & 
         \includegraphics[width=#8\textwidth]{fig/appendix/qa#2/kodak#1_bpg_crop_0.png} & 
         \includegraphics[width=#8\textwidth]{fig/appendix/qa#2/kodak#1_jpeg_crop_0.png} \\
   
         \includegraphics[width=#8\textwidth]{fig/appendix/qa#2/kodak#1.png_input_crop_1.png} & 
         \includegraphics[width=#8\textwidth]{fig/appendix/qa#2/kodak#1.png_recon_crop_1.png} & 
         \includegraphics[width=#8\textwidth]{fig/appendix/qa#2/kodak#1_ms_crop_1.png} & 
         \includegraphics[width=#8\textwidth]{fig/appendix/qa#2/kodak#1_bpg_crop_1.png} & 
         \includegraphics[width=#8\textwidth]{fig/appendix/qa#2/kodak#1_jpeg_crop_1.png} \\
       \end{tabular}
    \caption{Compression results including the source image, our result, M\&S Hyperprior model~\cite{minnen2018joint}, BPG (4:4:4) and JPEG (4:2:0) on Kodak #1 image.
            Each number in parentheses indicates bpp/PSNR (dB)/MS-SSIM of the reconstructed image.
            We adapt the compression rate of our model to that of M\&S Hyperprior model by adjusting the value of the uniform quality map.
            Our model outperforms all other methods in terms of the visual quality and PSNR/MS-SSIM metrics at similar bit rates.
            }
    \label{fig:appendix_kodak_#1}
   \end{figure*}
}

\qkodak{5}{2}{(0.2406 / 25.46 / 0.9304)}{(0.2406 / 24.85 / 0.9272)}{(0.2405 / 24.85 / 0.9173)}{(0.2513 / 21.34 / 0.8398)}{.43}{.158}
\qkodak{14}{4}{(0.1579 / 26.49 / 0.9013)}{(0.1579 / 26.02 / 0.8975)}{(0.1570 / 25.90 / 0.8812)}{(0.1732 / 22.49 / 0.7856)}{.43}{.158}
\qkodakv{19}{3}{(0.1061 / 28.69 / 0.9127)}{(0.1061 / 27.83 / 0.9067)}{(0.1061 / 28.41 / 0.9029)}{(0.1128 / 21.64 / 0.7177)}{.279}{.158}
\qkodak{22}{1}{(0.1772 / 29.18 / 0.9215)}{(0.1772 / 28.92 / 0.9168)}{(0.1855 / 29.03 / 0.9143)}{(0.1901 / 25.89 / 0.8336)}{.43}{.158}